\documentclass[%
 floatfix,
 reprint,
 twocolumn,
superscriptaddress,
showpacs,preprintnumbers,
 amsmath,amssymb,
 aps,
prd,
]{revtex4-1}

\usepackage{graphicx}  
\usepackage{dcolumn}   
\usepackage{bm}        
\usepackage{amssymb}   
\usepackage{comment} 
\usepackage{color}
\usepackage[percent]{overpic}
\usepackage{amsmath}
\usepackage{footmisc}

\usepackage[normalem]{ulem}

\begin{document}
\widetext


\title{Measurement of the  ($\pi^-$, Ar) total hadronic cross section  at the LArIAT experiment.}

\collaboration{LArIAT Collaboration}
\thanks{lariat\_authors@fnal.gov}\noaffiliation

\newcommand{\ABC}{Universidade Federal do ABC, Santo Andr\'{e}, SP 09210-580, Brasil}
\newcommand{\Alfenas}{Universidade Federal de Alfenas, Po\c{c}os de Caldas, MG 37715-400, Brasil}
\newcommand{\Boston}{Boston University, Boston, MA 02215, USA}
\newcommand{\Campinas}{Universidade Estadual de Campinas, Campinas, SP 13083-859, Brasil}
\newcommand{\Chiba}{Chiba University, Chiba 260, Japan}
\newcommand{\Chicago}{University of Chicago, Chicago, IL 60637, USA}
\newcommand{\Cincinnati}{University of Cincinnati, Cincinnati, OH 45221, USA}
\newcommand{\Edinburgh}{University of Edinburgh, Edinburgh EH9 3FD, UK}
\newcommand{\FNAL}{Fermi National Accelerator Laboratory, Batavia, IL 60510, USA}
\newcommand{\Goia}{Universidade Federal de Goi\'{a}s, Goi\'{a}s, CEP 74690-900, Brasil}
\newcommand{\KEK}{High Energy Accelerator Research Organization (KEK), Tsukuba 305-0801, Japan}
\newcommand{\Louisiana}{Louisiana State University, Baton Rouge, LA 70803, USA}
\newcommand{\Manchester}{University of Manchester, Manchester M13 9PL, UK} 
\newcommand{\Michigan}{Michigan State University, East Lansing, MI 48824, USA}
\newcommand{\Minnesota}{University of Minnesota, Duluth, Duluth, MN 55812, USA}
\newcommand{\NCBJ}{National Centre for Nuclear Research (NCBJ), Otwock 05-400, Poland}
\newcommand{\Syracuse}{Syracuse University, Syracuse, NY 13244, USA}
\newcommand{\Arlington}{University of Texas at Arlington, Arlington, TX 76019, USA}
\newcommand{\Austin}{University of Texas at Austin, Austin, TX 78712, USA}
\newcommand{\London}{University College London, London WC1E 6BT, UK}
\newcommand{\William}{William \& Mary, Williamsburg, VA 23187, USA}
\newcommand{\Yale}{Yale University, New Haven, CT 06520, USA}

\newcommand{\gramelliniThanks}{Present address: Fermi National Accelerator Laboratory, Batavia, IL 60510, USA}
\newcommand{\hoThanks}{Present address: Harvard University, Boston, MA 02138, USA}
\newcommand{\foremanThanks}{Present address: Illinois Institute of Technology, Chicago, IL 60616, USA}
\newcommand{\garciagamezThanks}{Present address: University of Granada, 18010 Granada, Spain}
\newcommand{\hillThanks}{Present address: Chiba University, Chiba 263-8522, Japan}
\newcommand{\kryczynskiThanks}{also Institute of Nuclear Physics PAN, 31-342 Krak\'{o}w, Poland}
\newcommand{\nunesThanks}{Present address: Syracuse University, Syracuse, NY 13244, USA}
\newcommand{\nutiniThanks}{also Istituto Nazionale di Fisica Nucleare, Italy and Gran Sasso Science Institute}
\newcommand{\olivierThanks}{Present address: University of Rochester, Rochester, NY 14627, USA}
\newcommand{\rebelThanks}{also University of Wisconsin-Madison, Madison, WI 53706, USA}
\newcommand{\rosslonerganThanks}{also Durham University, Durham DH1 3LE, UK}
\newcommand{\soubasisThanks}{Present address: Vanderbilt University, Nashville, TN 37235, USA}
\newcommand{\spagliardiThanks}{Present address: University of Oxford, Oxford OX1 3PJ, UK}

\affiliation{\ABC}
\affiliation{\Alfenas}
\affiliation{\Boston}
\affiliation{\Campinas}
\affiliation{\Chiba}
\affiliation{\Chicago}
\affiliation{\Cincinnati}
\affiliation{\Edinburgh}
\affiliation{\FNAL}
\affiliation{\Goia}
\affiliation{\KEK}
\affiliation{\Louisiana}
\affiliation{\Manchester} 
\affiliation{\Michigan}
\affiliation{\Minnesota}
\affiliation{\NCBJ}
\affiliation{\Syracuse}
\affiliation{\Arlington}
\affiliation{\Austin}
\affiliation{\London}
\affiliation{\William}
\affiliation{\Yale}

\author{E.~Gramellini}\thanks{\gramelliniThanks}\affiliation{\Yale}
\author{J.~Ho}\thanks{\hoThanks}\affiliation{\Chicago}
\author{R.~Acciarri}\affiliation{\FNAL}
\author{C.~Adams}\affiliation{\Yale}
\author{J.~Asaadi}\affiliation{\Arlington}
\author{M.~Backfish}\affiliation{\FNAL}
\author{W.~Badgett}\affiliation{\FNAL}
\author{B.~Baller}\affiliation{\FNAL}
\author{V.~Basque}\affiliation{\Manchester}
\author{O.~Benevides~Rodrigues}\affiliation{\Goia}\affiliation{\Syracuse}
\author{F.~d.~M.~Blaszczyk}\affiliation{\Boston}
\author{R.~Bouabid}\affiliation{\Chicago}
\author{C.~Bromberg}\affiliation{\Michigan}
\author{R.~Carey}\affiliation{\Boston}
\author{R.~Castillo~Fernandez}\affiliation{\FNAL}
\author{F.~Cavanna}\affiliation{\FNAL}\affiliation{\Yale}
\author{J.~I.~Cevallos~Aleman}\affiliation{\Chicago}
\author{A.~Chatterjee}\affiliation{\Arlington}
\author{P.~Dedin}\affiliation{\Campinas}
\author{M.~V.~dos~Santos}\affiliation{\Alfenas}
\author{D.~Edmunds}\affiliation{\Michigan}
\author{C.~Escobar}\affiliation{\FNAL}
\author{J.~Esquivel}\thanks{\gramelliniThanks}\affiliation{\Syracuse}
\author{J.~J.~Evans}\affiliation{\Manchester} 
\author{A.~Falcone}\affiliation{\Arlington}
\author{W.~Flanagan}\affiliation{\Austin}
\author{B.~T.~Fleming}\affiliation{\Yale}
\author{W.~Foreman}\thanks{\foremanThanks}\affiliation{\Chicago}
\author{D.~Garcia-Gamez}\thanks{\garciagamezThanks}\affiliation{\Manchester} 
\author{D.~Gastler}\affiliation{\Boston}
\author{T.~Ghosh}\affiliation{\Goia}
\author{R.~A.~Gomes}\affiliation{\Goia}
\author{R.~Gran}\affiliation{\Minnesota}
\author{D.~R.~Gratieri}\affiliation{\Campinas}
\author{P.~Guzowski}\affiliation{\Manchester} 
\author{A.~Hahn}\affiliation{\FNAL}
\author{P.~Hamilton}\affiliation{\Syracuse}
\author{C.~Hill}\affiliation{\Manchester} 
\author{A.~Holin}\affiliation{\London}
\author{J.~Hugon}\affiliation{\Louisiana}
\author{E.~Iwai}\affiliation{\KEK}
\author{D.~Jensen}\affiliation{\FNAL}
\author{R.~A.~Johnson}\affiliation{\Cincinnati}
\author{H.~Kawai}\affiliation{\Chiba}
\author{E.~Kearns}\affiliation{\Boston}
\author{E.~Kemp}\affiliation{\Campinas}
\author{M.~Kirby}\affiliation{\FNAL}
\author{T.~Kobilarcik}\affiliation{\FNAL}
\author{M.~Kordosky}\affiliation{\William}
\author{P.~Kryczy\'nski}\thanks{\kryczynskiThanks}\affiliation{\FNAL}
\author{K.~Lang}\affiliation{\Austin}
\author{R.~Linehan}\affiliation{\Boston}
\author{S.~Lockwitz}\affiliation{\FNAL}
\author{X.~Luo}\affiliation{\Yale}
\author{A.~A.~B.~Machado}\affiliation{\Campinas}
\author{A.~Marchionni}\affiliation{\FNAL}
\author{T.~Maruyama}\affiliation{\KEK}
\author{L.~Mendes~Santos}\affiliation{\Campinas}
\author{W.~Metcalf}\affiliation{\Louisiana}
\author{C.~A.~Moura}\affiliation{\ABC}
\author{R.~Nichol}\affiliation{\London}
\author{I.~Nutini}\thanks{\nutiniThanks}\affiliation{\FNAL}
\author{A.~Olivier}\thanks{\olivierThanks}\affiliation{\Louisiana}
\author{O.~Palamara}\affiliation{\FNAL}\affiliation{\Yale}
\author{J.~Paley}\affiliation{\FNAL}
\author{I.~Parmaksiz}\affiliation{\Arlington}
\author{B.~Passarelli~Gelli}\affiliation{\Campinas}
\author{L.~Paulucci}\affiliation{\ABC}
\author{D.~Phan}\affiliation{\Austin}
\author{G.~Pulliam}\affiliation{\Syracuse}
\author{J.~L.~Raaf}\affiliation{\FNAL}
\author{B.~Rebel}\thanks{\rebelThanks}\affiliation{\FNAL}
\author{M.~Reggiani~Guzzo}\affiliation{\Campinas}
\author{M.~Ross-Lonergan}\thanks{\rosslonerganThanks}\affiliation{\FNAL}
\author{D.~W.~Schmitz}\affiliation{\Chicago}
\author{E.~Segreto}\affiliation{\Campinas}
\author{D.~Shooltz}\affiliation{\Michigan}
\author{D.~Smith}\affiliation{\Boston}
\author{M.~Soares~Nunes}\thanks{\nunesThanks}\affiliation{\Campinas}
\author{M.~Soderberg}\affiliation{\Syracuse}
\author{B.~Soubasis}\thanks{\soubasisThanks}\affiliation{\Austin}
\author{F.~Spagliardi}\thanks{\spagliardiThanks}\affiliation{\Manchester} 
\author{J.~M.~St.~John}\thanks{\gramelliniThanks}\affiliation{\Cincinnati}
\author{M.~Stancari}\affiliation{\FNAL}
\author{D.~Stefan}\affiliation{\NCBJ}
\author{M.~Stephens}\affiliation{\William}
\author{R.~Sulej}\affiliation{\NCBJ}
\author{A.~M.~Szelc}\affiliation{\Edinburgh} 
\author{M.~Tabata}\affiliation{\Chiba}
\author{D.~Totani}\affiliation{\FNAL}
\author{M.~Tzanov}\affiliation{\Louisiana}
\author{D.~Walker}\affiliation{\Louisiana}
\author{H.~Wenzel}\affiliation{\FNAL}
\author{Z.~Williams}\affiliation{\Arlington}
\author{T.~Yang}\affiliation{\FNAL}
\author{J.~Yu}\affiliation{\Arlington}
\author{S.~Zhang}\affiliation{\Boston}
\author{J.~Zhu}\affiliation{\FNAL}

\date{\today}

\begin{abstract}
We present the first measurement of the negative pion total hadronic cross section on argon in a restricted phase space, which we performed at the Liquid Argon In A Testbeam (LArIAT)  experiment. All hadronic reaction channels, as well as hadronic elastic interactions with scattering angle greater than 5~degrees are included. The pions have a kinetic energies in the range 100-700~MeV and are produced by a beam of charged particles impinging on a solid target at the Fermilab Test Beam Facility. LArIAT employs a 0.24~ton active mass Liquid Argon Time Projection Chamber (LArTPC) to measure the pion hadronic interactions. For this measurement, LArIAT has developed the ``thin slice method", a new technique to measure cross sections with LArTPCs.  While moderately higher, our measurement of the  ($\pi^-$,Ar) total hadronic cross section is generally in agreement with the Geant4 prediction.
\end{abstract}

\pacs{14.40.-n	Mesons,  14.40.Be	Light mesons (S=C=B=0), 13.75.-n	Hadron-induced low- and intermediate-energy reactions and scattering (energy $<$ 10 GeV)}
\maketitle
\section{\label{sec:Motivations}Motivations and Introduction}
This work presents the first measurement of the total hadronic cross section of negative pions on argon ($\pi^-$, Ar) in the energy range from 100 to 700~MeV performed by the Liquid Argon in a Testbeam (LArIAT) experiment. 
A renewed interest in hadronic cross sections, particularly cross sections on argon, has arisen within the modern neutrino experimental landscape due to the proliferation of Liquid Argon Time Projection Chamber (LArTPC) neutrino detectors such as DUNE~\cite{DUNE-CDRvol1} and the SBN program~\cite{sbn}, which benefit from the LArTPC technology fine-grained tracking, powerful calorimetry and particle identification capabilities. 

Neutrino experiments rely on the products of the neutrino interactions to identify and reconstruct neutrino flavor and energy. Pions ($\pi$) are common products of neutrino interactions, especially in resonant scattering, deep inelastic scattering, and coherent pion production, and can help identify the neutrino interaction type.

To date, the literature on the hadronic interactions of the particles produced in neutrino interactions on argon for energies relevant to the neutrino products, namely below  1~GeV, remains scarce. Measuring these interaction channels for pions on argon is of particular importance for neutrino experiments in order to model the behavior of the pion both inside the nuclear matter and also at the larger scale of traversing the detector medium. Neutrino event generators and detector simulation packages base their pion transportation in argon on interpolations of the historical cross section measurements of lighter and heavier nuclei. One of the primary goals of LArIAT's dedicated measurement program is to bridge this gap in data, thus reducing the uncertainties related to pion interactions on argon.

At the neutrino event generator level, assumptions in the nuclear modeling and interactions of hadrons inside the nucleus must be made in order to work backward from the products of the neutrino interaction to reconstruct the neutrino energy and flavor. 
Pions produced in a neutrino interaction will experience hadronic interactions while still within the target nucleus (referred to as final state interactions) as well as during propagation through the detector volume.  Processes such as pion absorption and pion charge exchange can greatly modify the topology of a neutrino interaction in the detector and potentially lead to modifications in the event classification.
Dedicated measurements of pion-argon hadronic cross sections are needed to constrain the modeling of both of these effects in experiments reconstructing neutrino interactions.  
The ability to reconstruct the details of pion interactions inside the detector is essential for modern argon neutrino experiments to achieve the design resolution for their key physics measurements~\cite{Friedland:2020cdp}. 

This work is organized in several sections. To start, we briefly review previous pion hadronic cross section measurements on lighter and heavier nuclei and define the total pion hadronic cross section that is measured in LArIAT. In the next section, we present a brief overview of the LArIAT experimental setup, as well as an overview of the reconstruction and simulation techniques employed to identify ($\pi^-$, Ar) interactions. Next, we present a discussion of the ``thin-slice'' method, a new  method to measure cross sections on argon in a LArTPC, developed by the LArIAT collaboration.  Finally the ($\pi^-$, Ar) cross section measurement and its associated uncertainties are presented, followed by a brief discussion of the result.

\subsection{Historical Measurements of Pion Hadronic Cross Sections: Lighter and Heavier Nuclei}
Several experiments using pion beams have studied the hadronic interaction of pions on light and heavy nuclei, such as He, Li, C, S, Fe, and Pb~\cite{Wilkin:1973xd, Clough1974,PhysRevC.14.635,PhysRevC.23.2173}.  Most historical measurements of the pion hadronic cross sections were performed using ``thin targets'' with thicknesses much smaller than the typical interaction length in the material. At their core, these experiments are done by impinging a beam of pions of a known flux on a thin slab of material and recording the outgoing flux. Flux conservation allows retrieval of the interacting flux, and calculation of the cross section at a given beam energy. 

\begin{figure}
  \centering  
\includegraphics[width =0.5\textwidth]{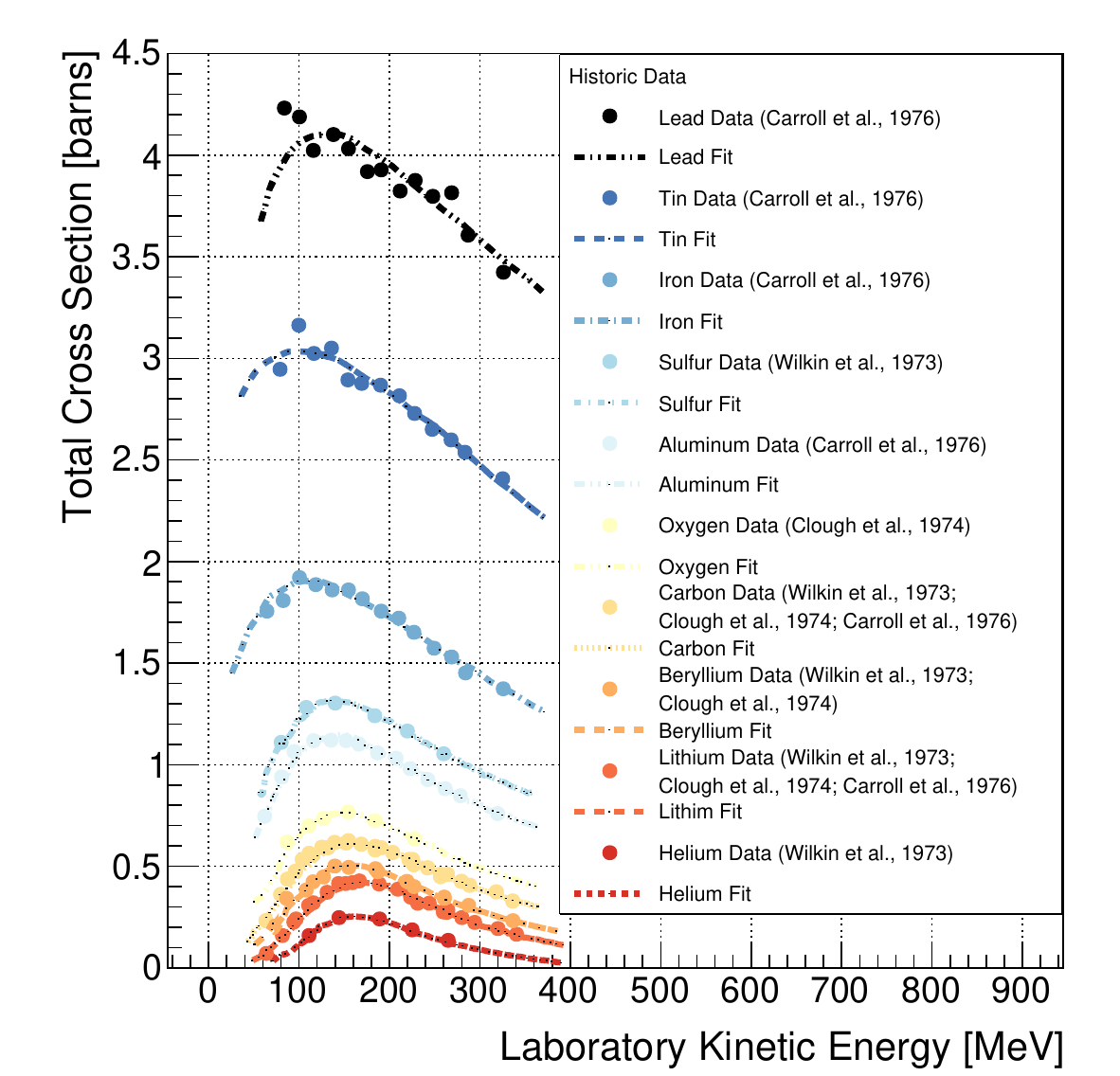}
\caption{ Historical positive pion scattering data for different light and heavy nuclei, reproduced from Reference~\cite{PhysRevC.14.635}.}
\label{fig:HistoricalData}
\end{figure}

Regardless of the nuclear medium, the shape of the pion-nucleus interaction cross section in the energy range accessible to  LArIAT  shows the distinct feature of a resonance. Indeed, a delta resonance ($\Delta$ 1232) is often produced in the pion-nucleon interaction, which subsequently decays inside the nucleus. Historical experimental results as the ones reported in Figure~\ref{fig:HistoricalData} for positive pions show a dependency of the delta resonance shape as a function of the nuclear mass number~\cite{PhysRevC.14.635}, typically parameterized by a Breit-Wigner function. The delta resonance shape becomes less pronounced and its peak shifts to lower energies as the nuclear mass number increases. This effect is due to the delta kinematics and its  propagation inside the nuclear medium, as well as to the nuclear potential on pions. Multiple scattering effects modify the resonance width, which is larger than the natural-decay width. This behavior makes interpolation of the inclusive cross section behavior from lighter nuclei to heavier nuclei subject to theoretical uncertainty and further motivates the necessity of such measurements on argon.

\subsection{\label{sec:Sign} $\pi^{-}$Ar Hadronic Interactions: Signal Signatures}
Strong hadronic interaction models~\cite{9780198520085,9780471779957} predict the pion interaction processes on argon in the hundreds of MeV energy range. The total hadronic ($\pi^-$, Ar)  cross section defines the probability of a single hadronic process on argon.  In measuring the total cross section ($\sigma_\text{Tot}$), we consider both the elastic interactions ($\sigma_\text{Elastic}$) and all hadronic reaction ($\sigma_\text{Reaction}$) channels, 
\begin{equation}
\sigma_\text{Tot} = \sigma_\text{Elastic}+ \sigma_\text{Reaction}.
\end{equation}
The hadronic reaction channel can be further subdivided into several exclusive channels with defined topologies:
\begin{equation}
\sigma_\text{Reaction} = \sigma_\text{Inelastic} + \sigma_\text{abs} + \sigma_\text{chex}+ \sigma_\text{$\pi$ prod}.
\end{equation}

\begin{figure}
  \centering  
\includegraphics[width =\columnwidth]{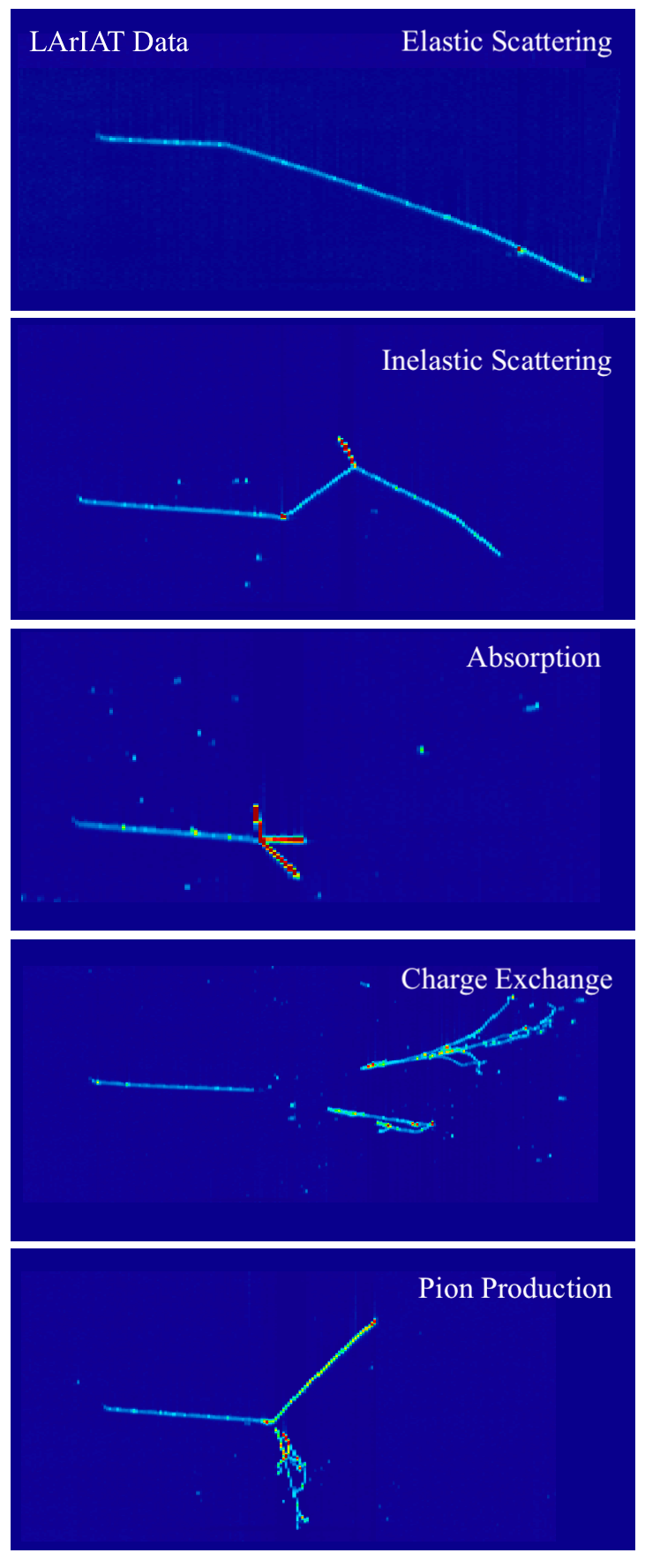}
\caption{Candidates for pion-argon interaction topologies as seen in the LArIAT data (from top to bottom): elastic scattering, inelastic scattering, pion absorption with emission of nucleons, charge exchange, and production of pions. The event displays show the raw signals in the time vs. wire space for the collection plane only, with beam particles entering from the left. The vertical and horizontal axes are not to the same scale in each image.}
\label{fig:PionsEvd}
\end{figure}

The term $\sigma_\text{Inelastic}$ defines hadronic inelastic interactions where the pion interacts with the nucleus with sufficient momentum transfer to cause nucleon knockout or to create a nuclear excited state with no nucleon emission, $\sigma_\text{abs}$ includes pion absorption on the nucleus, $\sigma_\text{chex}$ defines pion charge exchange where the incident charged pion converts to a neutral pion and the target nucleon undergoes a similar charge swap (e.g., $\pi^{-} p \rightarrow \pi^{0}n$), and $\sigma_\text{$\pi$ prod}$ includes pion production where the interaction energy of the incident pion is sufficient to produce additional pions in the collision. In this work, we account for all exclusive channels, regardless of their final state, in the total hadronic $\pi^{-}$-Ar interaction. For the elastic channel, we measure only interactions whose interaction angle -- defined as the angle of the scattered pion with respect to the incident pion direction -- is greater than 5~degrees. This phase space restriction is driven by the tracking algorithm efficiency and the objective of removing backgrounds from multiple Coulomb scattering (MCS). This is discussed further in section~\ref{sec:Tracking}.
Figure~\ref{fig:PionsEvd} shows examples of topologies for pion-argon hadronic interactions as they appear in the  LArIAT data.

\section{\label{sec:ExperimentalSetup}Experimental Setup}
The LArIAT experiment consists of a LArTPC deployed in a beam of charged particles at the Fermilab Test Beam Facility, situated in the  Meson Center beam line. During LArIAT's three seasons of data-taking, the experimental setup changed significantly in both its beamline and LArTPC components. The data used in this work were acquired during the 24 weeks of data-taking of the second season, which is referred to as Run~II~\cite{Acciarri_2020}. In the subsequent sections, we describe the beamline detectors and LArTPC configuration relevant to the cross section measurement. Following this discussion, we describe the essential components of the simulation, beamline event selection, and LArTPC event reconstruction used in this analysis.  The reference system adopted throughout the analysis is defined with respect to the LArTPC element. The coordinate origin is located at the front face of the LArTPC at middle height on the anonde plane; the $z$ direction correponds to the LArTPC main axis, pointing from the front face to the back face, the $y$ direction points against gravity and $x$ points from the anode to the cathode. 
We refer the reader to Ref.~\cite{Acciarri_2020} for a complete and detailed description of LArIAT's experimental setup.

\subsection{\label{sec:Beamline}Beamline} 
\begin{figure*}
\includegraphics[width=\textwidth,height=\textheight,keepaspectratio]{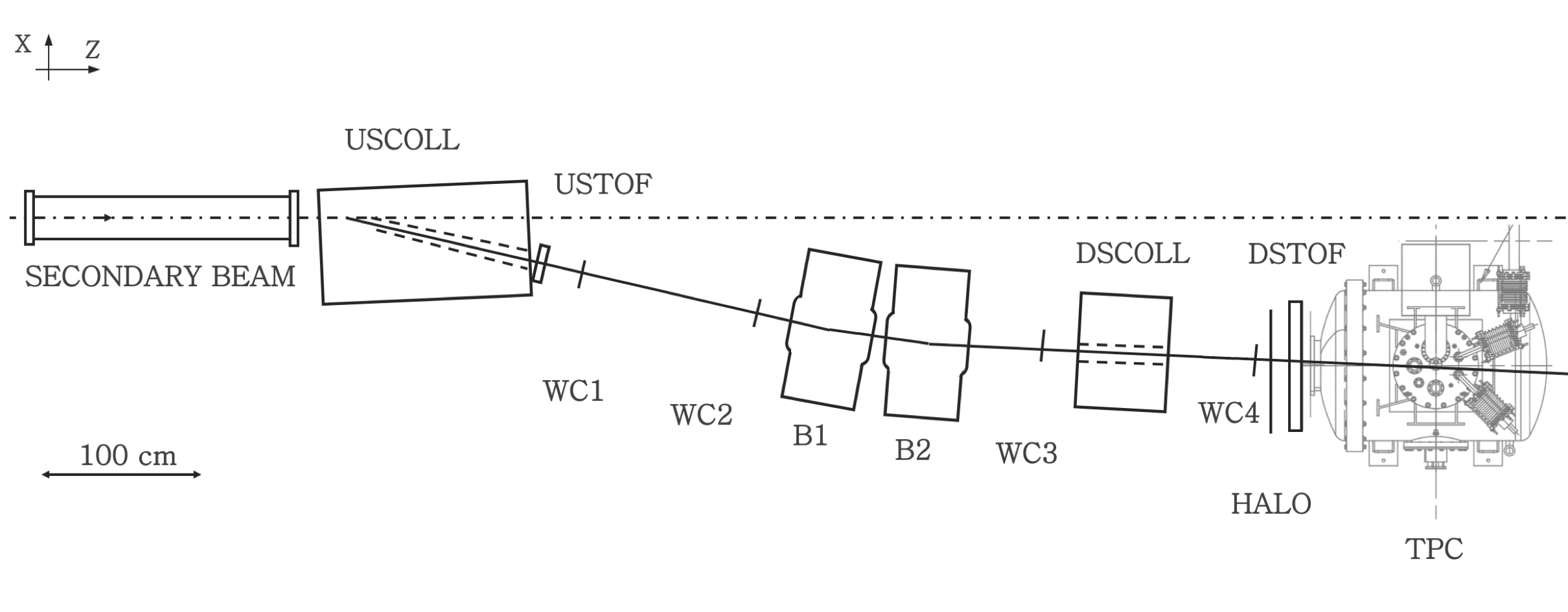}
\caption{Bird's eye view of the LArIAT tertiary beamline. USCOLL and DSCOLL represent the upstream and downstream collimators respectively; B1 and B2 represent the bending magnets; WC1, WC2, WC3 and WC4 are the multi-wire proportional chambers; USTOF and DSTOF represent the upstream and downstream time-of-flight scintillators; HALO represents a scintillator paddle with a central hole to enable vetoes of particles with trajectories incompatible with the beamline path; TPC shows the technical drawing of the cryostat which surrounds the liquid argon time projection chamber.}
\label{fig:beamlinebird}
\end{figure*}

LArIAT utilizes the Fermilab accelerator infrastructure in order to create a beam of charged particles.  The Fermilab accelerator complex  delivers a  primary beam of 120~GeV protons with variable intensity to the Meson Center beam line. This primary beam is focused onto a tungsten target to create a secondary beam, which is tuned such that its composition is mostly positive pions. For the data considered in this work, the tunable momentum peak of the secondary beam was fixed at 64~GeV/c; this configuration assured stable beam delivery at the LArIAT experimental hall (MC7).

\begin{figure}
\centering  
\includegraphics[width =\columnwidth]{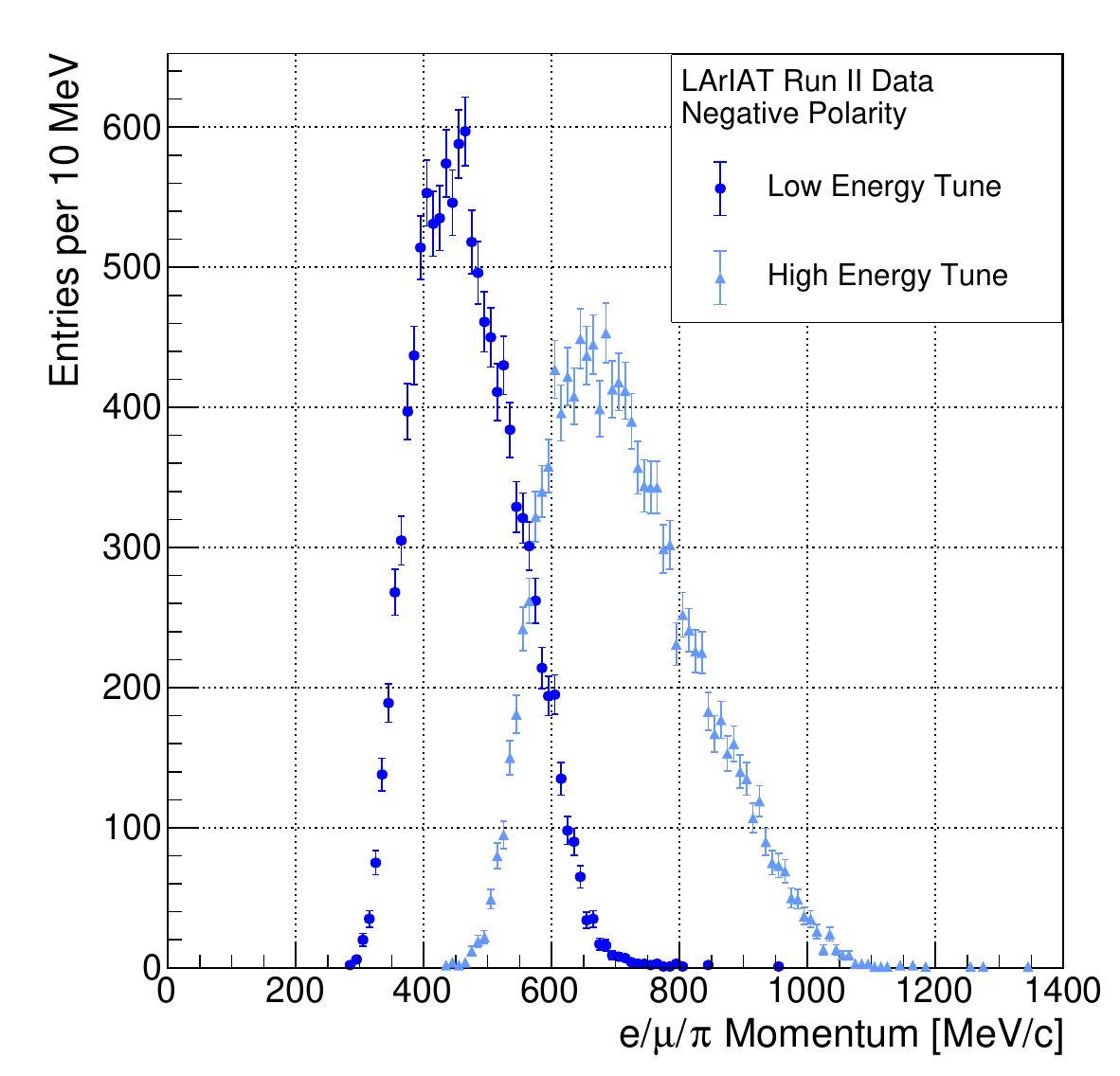}
\caption{Momentum spectrum in the LArIAT beamline  for Run~II negative polarity data, low energy tune in dark blue, high energy tune in light blue. }
\label{fig:momentum}
\end{figure}

At MC7, the secondary beam was focused onto a copper target within a steel collimator, creating and defining the LArIAT tertiary beam. The LArIAT tertiary beamline instrumentation identified the particle type and measured the momentum of the particles before they enter the LArIAT LArTPC.
Figure~\ref{fig:beamlinebird} shows a bird’s-eye view of the LArIAT tertiary beamline in MC7, consisting of two bending electromagnets (B1 and B2), a set of four wire chambers (WC1-WC4), and two time-of-flight scintillating paddles (USTOF and DSTOF).  The polarity of the magnets can be configured to charge-select the beam, and the magnetic field strength is tuned to select the range of momenta of particles steered toward the LArTPC. The combination of magnets and wire chambers forms the LArIAT spectrometer which measures the particles' momentum at the fourth wire chamber, $p_{\text{Beam}}$. 
Figure~\ref{fig:momentum} shows the distribution of  $p_{\text{Beam}}$ for the data used in this work; the momentum range of the two datasets used, one from a low momentum tune and the other from a higher momentum tune, spans from $\sim300$ to $\sim1100$~MeV/c.

A scintillator paddle (HALO) with a central hole was situated between DSTOF and the cryostat, just upstream of the LArTPC. Its purpose was to identify and reject particles that entered the LArTPC at a position that is outside the thin beam window in the cryostat wall. While information from this element is not used in this analysis, its passive material is considered when assessing the pion energy.

\begin{figure}
\centering  
\includegraphics[width =\columnwidth]{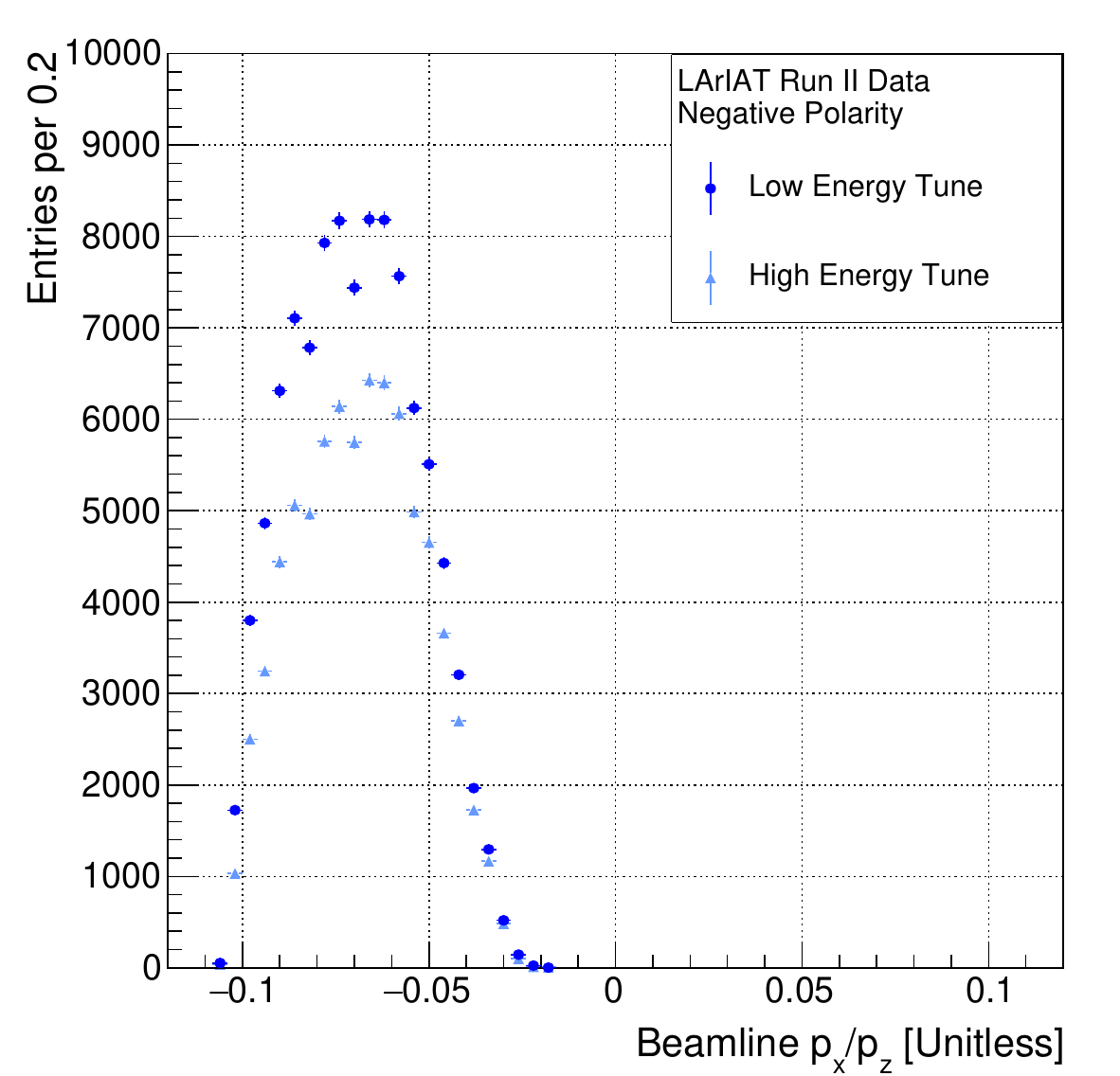}
\caption{Distribution of the p$_x$ to p$_z$ momentum ratio for electrons, muons, and pions in the LArIAT beamline  for Run~II negative polarity data, low energy tune in dark blue, high energy tune in light blue.  The $x$ component of the particles' momentum is small, since particles are mostly directed in the $z$ direction. The px/pz shows an offset of the mean relative to the beam axis due to the action of the bending magnets.}
\label{fig:momentumpx}
\end{figure}

\begin{figure}
\centering  
\includegraphics[width =\columnwidth]{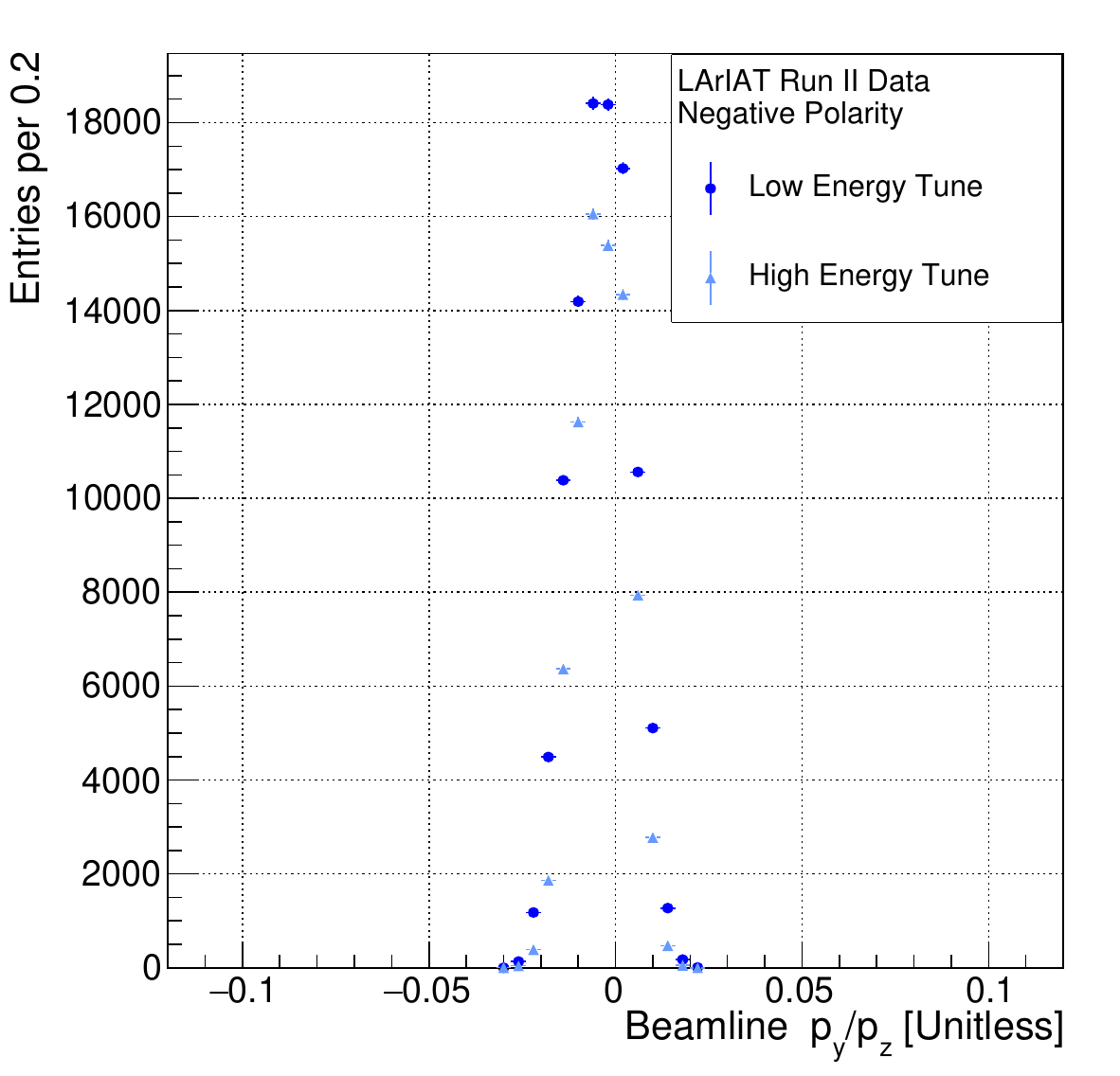}
\caption{Distribution of the p$_y$ to p$_z$ momentum ratio for electrons, muons, and pions in the LArIAT beamline  for Run~II negative polarity data, low energy tune in dark blue, high energy tune in light blue. The $y$ component of the particles' momentum is small, since particles are mostly directed in the $z$ direction. As expected given the absence of magnetic field on the vertical axis, the py/pz is symmetric around the beam center. }
\label{fig:momentumpy}
\end{figure}

\subsection{\label{sec:LArTPC}LArTPC}

The LArIAT LArTPC is a box of dimensions 47~cm (drift) by 40~cm (height) by 90~cm (length) with an applied electric field of 490~V/cm. Two instrumented read-out wire planes, one induction plane and one collection plane, along with an uninstrumented shield plane, form the LArIAT anode. LArIAT's induction and collection planes used in this analysis consist of 240 wires, each at 4~mm spacing. The wires are oriented at +/-$60^{\circ}$ from the horizontal direction, while the beam direction is oriented 3$^\circ$ off the LArTPC's long axis in the drift direction. Beamline particles enter the LArTPC roughly at the center of the front face leaving traces of ionization in the LArTPC. The ionization signals on the wires are then recorded by the LArIAT data acquisition system (DAQ), stored, and processed offline. LArIAT utilizes the LArSoft toolkit~\cite{LArSoft} for data acquisition, signal processing, event reconstruction, and LArTPC simulation. 

\section{Reconstruction and Simulation}


Given LArIAT's variety of sub-detectors, dedicated reconstruction and simulation techniques are needed to measure the physics quantities of interest. The next sections describe the methods used to extract information from the beamline detectors and from the LArTPC.

\subsection{\label{sec:BeamlineReco}Information from the Beamline}
Information from the beamline detectors used for physics analyses corresponds to three main components: the time-of-flight system, the bending magnets, and the wire chambers, described in greater detail in Ref.~\cite{Acciarri_2020}. 

The LArIAT time-of-flight (TOF) detector system consists of two scintillator paddles which bracket the beamline. The difference in the signal times of the upstream and downstream paddles is used to form the measurement of the TOF with a typical resolution of $\pm 1$ns. 

The bending magnet layout was based upon the MINERvA beam test, Fermilab T977~\cite{MinervaTestbeam}; together with the wire chamber system, they act as a particle spectrometer. The direction of the magnetic field in the center of the magnets is along the vertical, $y$-axis, pointing up or down depending on the polarity of the magnet currents. The 3D positions of the hits in the upstream wire chambers provide a straight trajectory before the bending magnets, while the positions of the hits in the downstream wire chambers provide the subsequent straight trajectory. A charged particle traversing the beamline bends in the $xz$-plane. The measurement of the transverse component of the particle's  momentum, $p_\text{xz}$, is performed using the bend plane angles, $\theta_{\textrm{US}}$ and $\theta_{\textrm{DS}}$, of the upstream and downstream trajectories. The particle's total momentum prior to entering the LArTPC, as well as its three separate spatial components, can then be calculated from $p_\text{xz}$ using the particle's downstream 3D trajectory.

In data, the reconstruction of beamline events begins by considering the activity in the wire chambers and the TOF to form a hypothesized particle trajectory. The reconstruction verifies that a particle's hypothesized trajectory through the wire chambers is plausible. Events are rejected if the trajectory from the wire chambers would cross impenetrable material in the beamline, such as the steel of the magnet. The horizontal and vertical components of the momentum are obtained using the positions of the hits in WC3 and WC4. In Figure ~\ref{fig:momentumpx}, the deflection of the magnets is apparent in the horizontal component’s offset of the mean relative to the beam center. The vertical component of the momentum, shown in Figure~\ref{fig:momentumpy}, is not affected by the magnet bending, and is therefore symmetrical with respect to the beam center at zero. 

\begin{figure}
  \centering  
\includegraphics[width =\columnwidth]{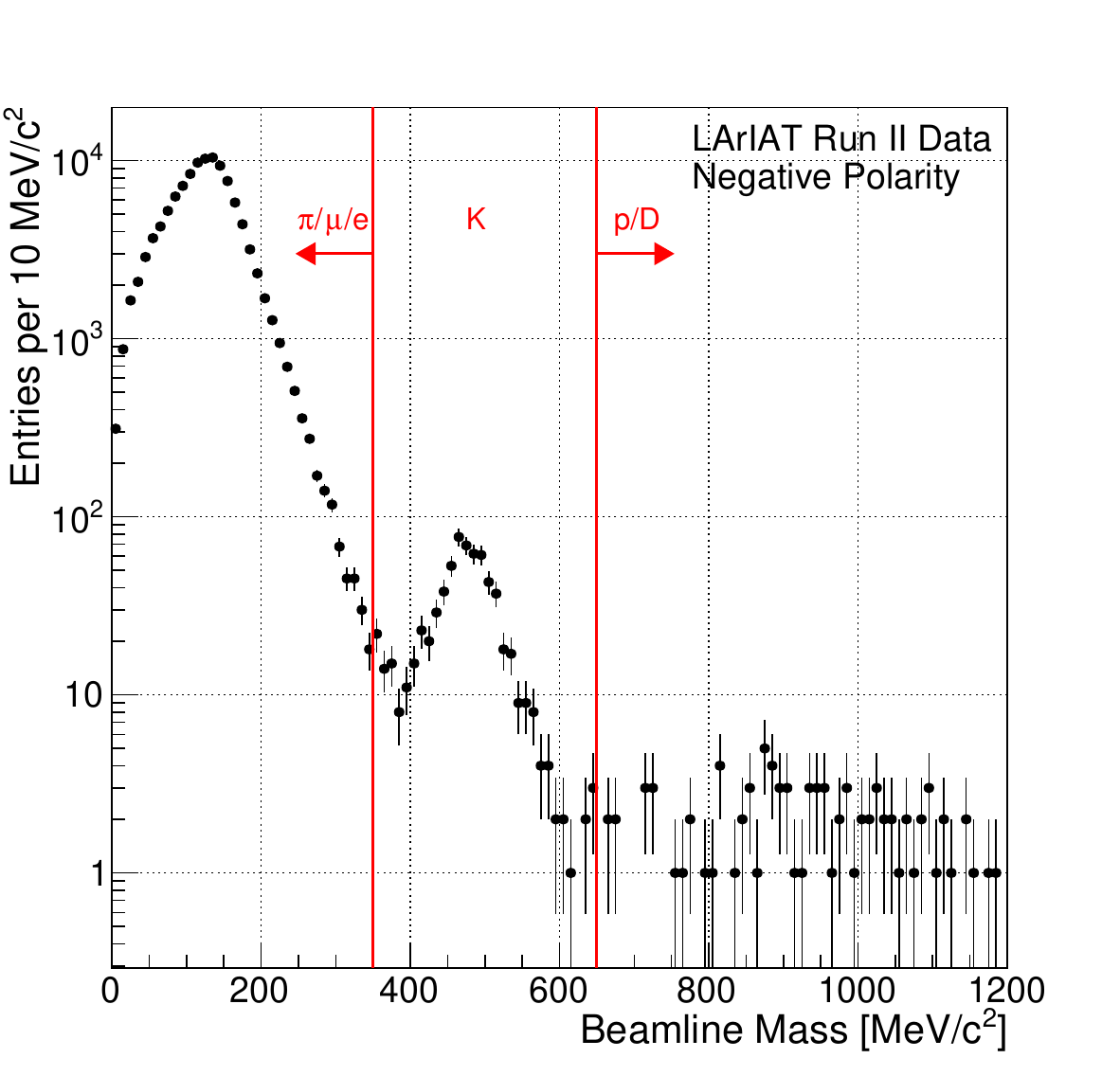}
\caption{Distribution of the beamline mass as calculated according to Equation~\ref{eq:mass} for the Run~II events reconstructed in the beamline, for negative polarity runs. The classification of the events into $\pi^-/ \mu^-/e^-$, K$^-$, or antiproton is based on these distributions, whose selection values are represented by the vertical red lines.}
\label{fig:mass}
\end{figure}

By combining the measurement of the particle's momentum with the measurement of the TOF, we calculate an invariant mass hypothesis, $m_{\text{Beam}}$, for the beamline particle as 
\begin{equation}
m_{\text{Beam}} = \frac{p_{\text{Beam}}}{c}\sqrt{\biggl(\frac{\text{TOF}\cdot c}{\ell}\biggr)^2 -1},
\label{eq:mass}
\end{equation}
where $c$ is the speed of light and $\ell$ is the length of the particle's trajectory between the time-of-flight paddles, typically 6.65~m. Due to the resolution of the time-of-flight measurement, an imaginary mass is computed for some events; for these events, $m_{\text{Beam}}$ corresponds to the absolute value of the mass in Equation~\ref{eq:mass}.
Figure~\ref{fig:mass} shows the distribution of the invariant mass for the entire negative polarity data set used in this analysis. We classify events into different particle hypotheses as follows according to the selection in Table~\ref{tab:Mass}.

\begin{table}
\caption{\label{tab:Mass} Beamline particle classification.}
\begin{ruledtabular}
\begin{tabular}{|l|llrl|}
ID&  &Selection &  &\\
\hline
 $\pi/\mu/e$ &                               & $m_{\text{Beam}}\leq$& 350~MeV/c$^2$&\\
 kaon           & 350~MeV/c$^2 <$ & $m_{\text{Beam}}\leq$& 650~MeV/c$^2$&\\
 antiproton   & 650~MeV/c$^2 <$ & $m_{\text{Beam}}\leq$& 3000~MeV/c$^2$&\\
\end{tabular}
\end{ruledtabular}
\end{table}

\subsection{\label{sec:Sim} MC Generation \& Simulation}
For this analysis, LArIAT employs a Monte Carlo (MC) simulation
to model particle interactions in the material directly upstream of the LArTPC as well as the response of the LArIAT LArTPC itself, implemented in the LArSoft framework. 

Simulated particles are generated at the fourth wire chamber and propagated downstream using a data-driven technique to match the properties of the real beam. The measurement of the beamline particles’ momenta and positions performed on data serve as the basis for the data-driven Monte Carlo (DDMC) event generation.

 
The DDMC simulation draws from the joint distribution of the momentum and position measurements in data using a 5-dimensional hit-or-miss sampling procedure. This sampling generates the simulated events  with the same momentum and position distributions as data (e.g., drawing from the distributions shown in Figures~\ref{fig:momentumpx} and~\ref{fig:momentumpy}), with the additional benefit of accounting for the correlations among the considered variables. Once completed, the LArSoft simulation uses the generated particle distributions to launch single particles from the location of the fourth wire chamber (100~cm upstream of the LArTPC). The particles are free to decay, interact, and deposit their energy along their path to the LArTPC according to the Geant4~\cite{Geant4} simulation and a detailed geometry of the intervening material between the fourth wire chamber and the LArTPC. When the DDMC particles arrive to the LArTPC, their simulated passage through matter is transformed into ionization signals which mimic the actual data. 

The DDMC samples are used in this analysis to propagate the estimated beamline background to the pion cross section, to calibrate the energy loss upstream of the LArTPC, and to study the tracking and the calorimetric performance in the LArTPC.

\subsection{\label{sec:TPCReco}LArTPC Reconstruction}
The particles from the beam data and from the DDMC propagate into the LArTPC. The processing and reconstruction of LArTPC signals is an active area of continuing development which spans from more traditional algorithmic approaches~\cite{Barker2011} to the use of machine learning tools~\cite{1748-0221-12-03-P03011}. Below, we summarize the processing and reconstruction chain of the LArTPC signals used in both LArIAT data and simulation to go from ionization-induced pulses on the sense wires to the construction of three-dimensional objects with associated calorimetric information. 

\subsubsection{Deconvolution}\label{sec:deconv}
As is typical in LArTPCs~\cite{Baller:2017ugz}, the first step in the LArTPC signal processing chain is deconvolution, which aims to remove the response of the readout electronics and to transform the induction and collection signals into a comparable set of waveforms on all planes presenting unipolar, approximately Gaussian-like pulses. Induction and collection plane signals have different field responses due to the different nature of the signals on these planes. The wires on the induction planes see the induced signal of the drifting ionization charge, which moves toward the wires and then passes by them without being collected, while the wires on the collection planes see the current derived from the charge entering the conductor material. Thus, prior to deconvolution, signals on the induction plane are bipolar pulses and signals on the collection plane are unipolar pulses. 

\subsubsection{Hit Reconstruction}\label{sec:HitReco}
The second stage of the signal processing is the reconstruction of wire signals which indicate an energy deposition in the detector, known as a ``hit''.  An algorithm scans the deconvolved LArTPC waveforms for each wire over the entire readout time window, searching for peaks above the waveform's baseline. These peaks are fit with a Gaussian function and the best fit parameters are stored, such as the peak time, height, width, and area under the Gaussian fit. The area of the Gaussian is proportional to the charge collected on the wire and the peak time is proportional to the coordinate in the drift direction where the ionization occurred. The event reconstruction chain uses the collection of hits to form more complex objects associated with the particles in the detector. 

\subsubsection{2D Clustering}\label{sec:Cluster}
Collections of hits, separately for each wire plane, are grouped together into objects known as ``clusters'', based on their topology. LArIAT identifies line-like objects 
through the use of the clustering package known as Trajectory-Cluster (TrajCluster)~\cite{Baller:2017ugz}. TrajCluster looks in the wire-time 2D projection for a collection of hits that can be described with a line-like trajectory. TrajCluster forms this collection by using the first two hits in the beam direction to form a ``seed trajectory''. The algorithm then subsequently steps through the other 2D hits to gather together points which belong to this assumed trajectory. Several factors determine whether a hit is added to the trajectory, including, but not limited to: the goodness of the fit of the single hit, the charge of the hit compared to the average charge and RMS of the hits already forming the trajectory, the goodness of trajectory fit with and without the hit addition, and the angle between the two lines formed by the collection of hits before and after the considered hit in the trajectory. The final product of this reconstruction stage is the collection of two-dimensional clusters on each wire plane. 

\subsubsection{3D Tracking}\label{sec:Tracking}
Collections of 2D clusters are matched between wire planes to form 3D objects. In this analysis, the 2D  clusters are used by the 3D tracking algorithms to form 3D tracks. This algorithm, first developed for the ICARUS collaboration~\cite{Antonello2013}, uses pairs of 2D clusters in the induction plane and collection plane which are close in time as a starting point to form a 3D track. It constructs a tentative 3D trajectory using the edges of the clusters. The algorithm then projects back the tentative trajectory onto the 2D planes and adjusts the parameters of the 3D track such that they minimize the distance between the projections and the track hits on all wire planes simultaneously.  Three-dimensional tracking can use multiple clusters in one plane, but it can never break them into smaller groups of hits. The final product of this reconstruction stage is the formation of three-dimensional objects (tracks) in the LArTPC. \\

For this analysis, we are primarily interested in reconstructing the tracks of the pion candidates from their entry point into the LArTPC's active volume up to their interaction point. Since this inclusive cross section measurement does not distinguish among the various hadronic interaction channels, reconstruction of the outgoing particle trajectories is not relevant. Thus, we focus on ``track reconstruction," which at LArIAT’s beam energies is generally associated with the presence of hadrons, as opposed to ``shower reconstruction," which is generally associated with electromagnetic activity from electrons or photons in the detector.
Since our signal definition for a pion interaction includes all hadronic channels, it is the end point of the primary pion track within the fiducial volume which identifies the interaction location, regardless of the final state topology.
The clustering algorithm was tuned to maximize the efficiency of finding all hadronic interactions~\cite{osti_1489387}.   This algorithm decides if an interaction vertex is found by considering 4 consecutive hits. The average distance between hits is 4.7~mm, which corresponds to $\sim$2 cm of argon. In 2~cm of argon, the expected MCS for the lower energy pions considered in the analysis is approximately 2~degrees. The reconstruction was shown to perform consistently in data and simulation.
By defining the signal to only account for elastic hadronic interactions above 5 degrees, we conservatively exclude the MCS irreducible background. This choice of signal definition also reflects our ability to efficiently identify hadronic interactions: with the reconstruction tune used in the analysis, the efficiency of finding a kink below 5~degrees is significantly lower than the rest of the phase space.

\subsubsection{Calorimetry}\label{sec:Calo}
The last step in the event reconstruction chain is to assign calorimetric information to the track objects. Calorimetry is performed separately for each of the two wire planes. A multi-step procedure is needed to retrieve the energy deposited in the LArTPC from the charge seen by the wires. For each 2D hit associated with the 3D track object, the calorimetry algorithm calculates the charge seen on every wire by integrating the area underneath the Gaussian fit. This charge is then corrected to account for relative differences in wire response amplitudes, converted from units of ADC to electrons, and scaled up to correct for electron drift attenuation due to impurities. Charge recombination effects are then accounted for in order to derive the deposited energy.  Lastly, an overall calibration factor is applied, primarily based on energy deposits from minimum ionizing particles.   Thus the calorimetric energy deposit at every location on a given track has been assigned.  More details on the overall calibration are given in Reference~\cite{Acciarri_2020}.

\section{\label{sec:EventSelection}Pion Candidates Selection}

For each of the negative polarity Run~II beamline events, we select particles classified as $\pi/\mu/e$ by the magnetic spectrometer. We apply a series of additional event selection criteria to isolate the pions from these other light particles as much as possible.

Pile-up (multiple beamline particles entering the LArTPC at the same time) and beam halo (additional particles entering the LArTPC from points outside the cryostat's beam window) can affect the accuracy of the LArTPC track reconstruction. To mitigate these effects, we place a requirement on the number of tracks entering the LArTPC: events must have four or fewer tracks in the first 14~cm of the LArTPC active volume. 

We project the reconstructed beamline track direction and position to the front face of the LArTPC in order to match the beamline pion candidate with a reconstructed LArTPC track. A match is found when the projection of the beamline track is within 4 cm of a reconstructed LArTPC track, and the angle between the beamline and LArTPC track is less than 8~degrees.
Events are required to have one and only one beamline-to-LArTPC match.

\begin{figure}
\includegraphics[width = \columnwidth]{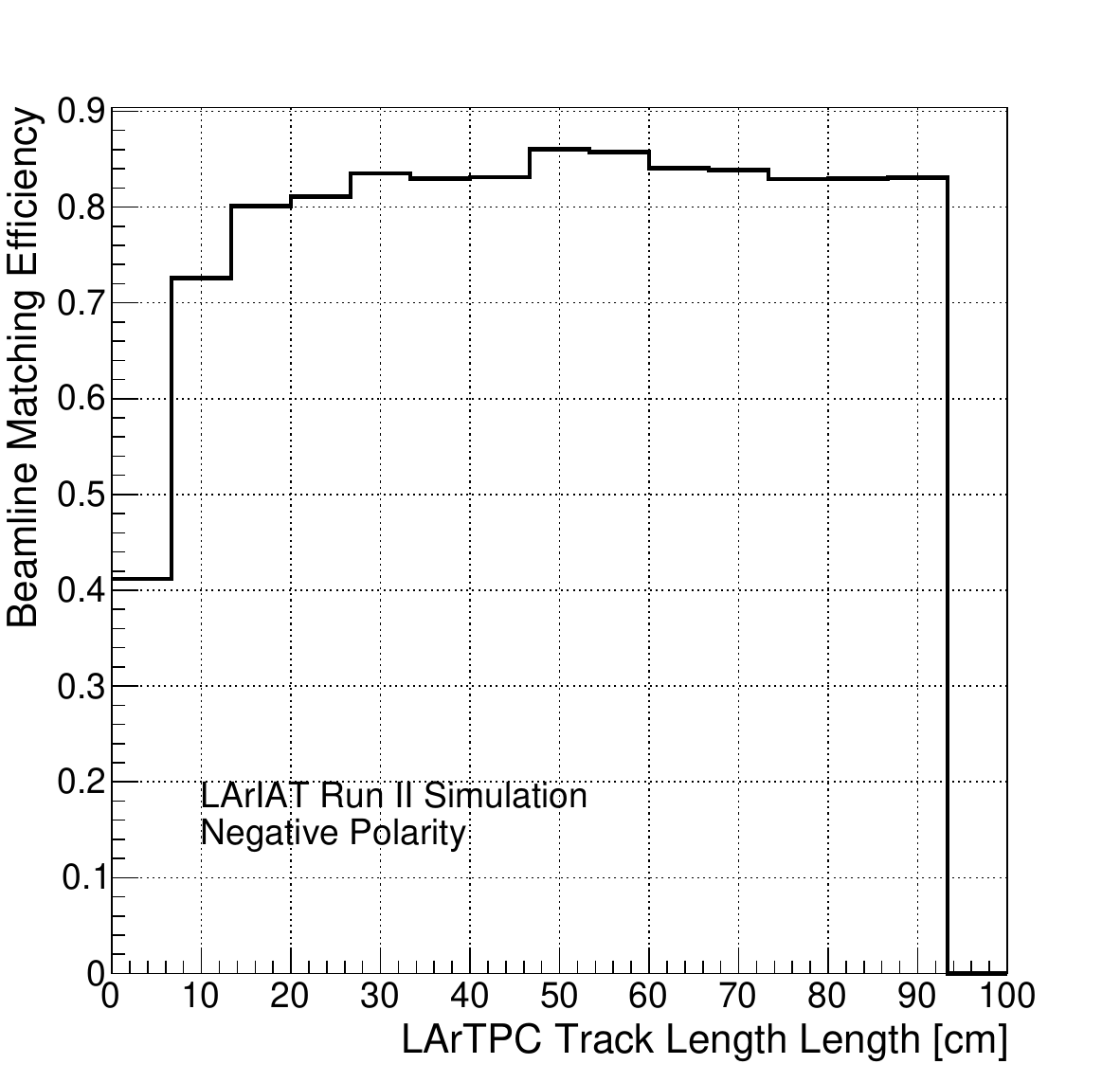}
\caption{\label{fig:eff} Beamline matching efficiency as a function of the reconstructed LArTPC track length. }
\end{figure}

The efficiency of matching beamline tracks to LArTPC tracks has some dependence on the length of the reconstructed LArTPC track, shown in Figure~\ref{fig:eff}, which has the potential to bias the final cross section measurement. To minimize bias, we define a LArTPC fiducial volume in which the beamline-to-LArTPC track-matching efficiency is invariant with respect to the reconstructed track length. Specifically we consider only events whose matched track penetrates to at least 30~cm from the front face of the LArTPC.  Any track which extends to more than 80~cm from the front face we label as ``exiting". Only the portions of the tracks  within the fiducial volume (in the 30~cm and 80~cm range) are used in the analysis. This LArTPC fiducialization also carries the benefit of eliminating much of the electron content.

The majority of the remaining electron contamination is eliminated by placing requirements on the reconstructed LArTPC tracks. We construct a ``shower filter" by leveraging the topological difference between the track-like signal generated by pions and the electromagnetic showers generated by electrons. 
When the track reconstruction is applied to electromagnetic activity in the LArTPC, the result is the reconstruction of numerous short-length tracks; if more than 5 tracks shorter than 10~cm are present near the matched LArTPC track, the event is classified as an electron and is rejected. After all of the selection requirements are applied, the electron content is $1 \pm 1\%$ for both the low and high energy tune, as reported in Table~\ref{tab:MCafterCutContaminants}.

These selection criteria unfortunately cannot eliminate the muon content in the beamline. Muon contamination is estimated through geometrical and kinematic considerations, exploiting the fact that  muons are present in the beamline only as a product of pion decays.  The pion candidates used in this analysis require a beamline trigger (i.e., in-time activity in all four wire chambers and TOF) and a matched reconstructed LArTPC track, which places geometrical constraints on the decay angle of muons relative to the direction of their parent pions in the lab frame. The beamline geometry, taken together with relativistic particle considerations, results in a conservative estimate of muon content in the beamline: $6.1 \pm 1.6 \%$ for the low energy tune and $7.8 \pm 1.4 \%$ for the high energy tune, as reported in Table~\ref{tab:MCafterCutContaminants}.

Starting from 158396 beamline triggers,  24534 beamline pion candidates remain in the Run~II negative polarity data sample after we apply the selection criteria. 

We feed beamline candidates to the ``thin slice method" machinery, discussed in the next section, in order to measure the negative pion total hadronic cross section. 

\begin{table}
\caption{\label{tab:MCafterCutContaminants}Expected beamline composition for the selected events in the two energy tunes.}
\begin{ruledtabular}
\begin{tabular}{ l | c | c | c || c | c | c |}
 &  \multicolumn{3}{c||}{Low Energy Tune} & \multicolumn{3}{c|}{High Energy Tune }\\
& MC $\pi$  & MC  $ \mu$ & MC  $e$ & MC  $\pi$ & MC  $\mu$ & MC  $e$ \\
\hline
&  &  &  & & &\\  
Expected&  92.9\%   & 6.1\%    & 1.0\%   & 91.2\%	& 7.8\%  & 1.0\%\\
Composition              &                      &$\pm$                       &$\pm$                   &                       &$\pm$                        &$\pm$\\  
in XS sample &                      &1.6\%&          1.0\% &                   &                 1.4\%&    1.0\%\\  
\end{tabular}
\end{ruledtabular}
\end{table}

\section{\label{sec:ThinSliceMethod}{Thin Slice Method}}
Pion hadronic cross sections have historically been measured  using ``thin targets'', where the thickness of the target is much smaller than the typical interaction length in the material. While the LArIAT LArTPC as a whole fails this constraint, its fine-grained tracking allows for the development of a methodology which still leverages the conditions of thin target experiments.

The pion interaction probability, $P_{\text{Int}}$, for an argon target of thickness $\delta X$  is related to the interaction cross section $\sigma_{\text{TOT}}$ by 
\begin{equation}
P_{\text{Int}} = 1- e^{-\sigma_{\text{TOT}}\text{ } n \text{ }\delta X},
\label{eq:thinTargetXS}
\end{equation}
where $n$ is the density of the target centers. The density of target centers is defined as $n=\frac{\rho N_{A} }{m_A}$, with argon mass density ($\rho$), Avogadro's number ($N_{A}$) and argon molar mass ($m_A$).  The probability of pion interaction on the target, $P_{\text{Int}}$, can be statistically estimated as the ratio of the number of pions interacting in the thin target, $N_{\text{Int}}$ (interacting flux), and the number of incident pions, $N_{\text{Inc}}$ (incident flux). 
If the interaction length is significantly longer than the thickness of the target, i.e $\lambda_{\text{Int}} \gg \delta X$, we can assume that the target centers are uniformly distributed in the material and that no center of interaction sits in front of another. This is typically known as the thin target approximation. 

In this approximation, it is possible to find a simple proportional relationship between the cross section and the interaction probability
from the first term of a Taylor expansion of the exponential function:
 \begin{equation}
 \sigma_{\text{TOT}}  = \frac{1}{n \text{ }\delta X}\frac{N_{\text{Int}}}{N_{\text{Inc}}}.
\label{eq:thinTargetXSSolved}
\end{equation}

\begin{figure}
\includegraphics[width = \columnwidth]{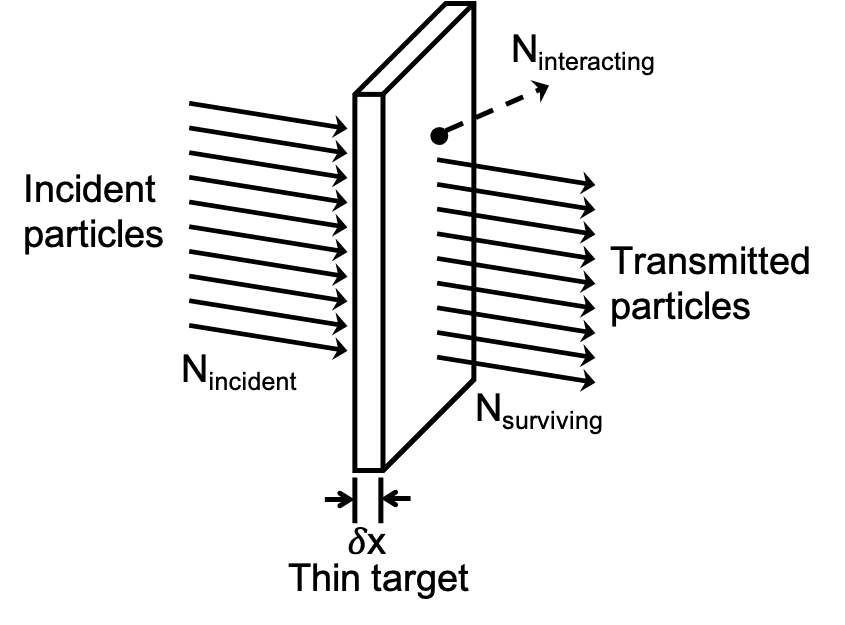}
\includegraphics[width = \columnwidth]{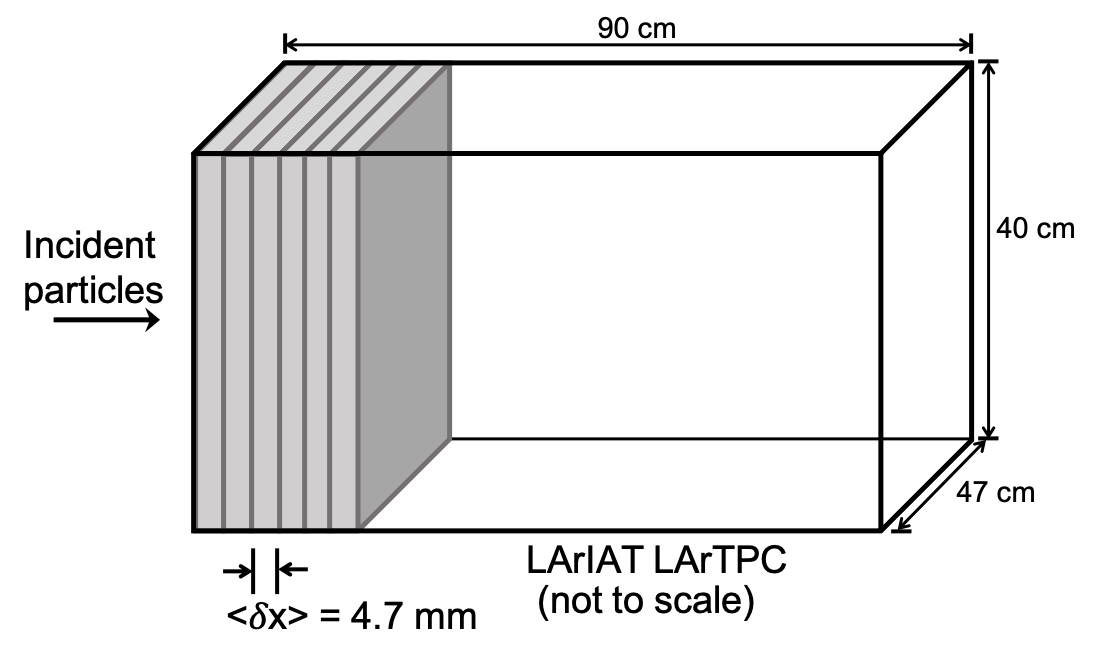}
\caption{\label{fig:ThinSlice} Graphical representation of the thin slice method and its implementation in the LArIAT LArTPC. The average slice thickness is determined by the wire angle and spacing in the readout wire planes ($\pm 60^{\circ}$ from vertical, 4~mm pitch). }
\end{figure}

Since the interaction length of pions in liquid argon is of the order of $\sim 50$~cm, the LArIAT LArTPC, with a length of 90~cm, does not represent a thin target. However, the granularity of the LArTPC allows us to measure the kinetic energy of each pion approximately every 4.7~mm along its trajectory in the detector's active volume. We may thus treat the argon volume as a sequence of many adjacent thin targets, as illustrated in Figure~\ref{fig:ThinSlice}, recovering the thin target approximation in each argon slice. 

Each slice of argon, {\emph{j}}, represents an independent thin target experiment for which the incoming pion has a kinetic energy of $E^{kin}_j$, given by the equation:

\begin{equation}
\begin{split}
 E_{j}^\text{kin}  & = E^\text{kin}_{\text{Initial}} -  \sum_{n < j} E_{\text{Dep},n}.
\end{split}
\label{eq:KEj}
\end{equation}
where $E^\text{kin}_{\text{Initial}}$ represents the initial kinetic energy the particle had upon entering the LArTPC, and $E_{\text{Dep},n}$ is the energy deposited in slice $n$. 

The pion loses kinetic energy in each traversed slice via ionization. Therefore, the same pion will contribute to subsequent independent thin target experiments at progressively decreasing energies the further it travels in the LArTPC.  We apply the cross section calculation from Equation~\ref{eq:thinTargetXSSolved} in bins of kinetic energy: when a pion of kinetic energy $E^{kin}_j$ enters a slice, it contributes to the incident flux at the energy bin corresponding to $E^{kin}_j$.  Within the $j^{th}$ slice, the pion may or may not undergo a hadronic interaction. If it does, it also contributes to the interacting flux at the same energy bin. If the pion does not interact in the $j^{th}$ slice, it will enter the next slice and the evaluation of the fluxes is repeated for the new kinetic energy $E^\text{kin}_{j+1}$. The kinetic energy $E^\text{kin}_j$ is always greater than $E^\text{kin}_{j+1}$ while the pion is traversing the LArTPC. The contributions to the incident and interacting fluxes of all the beamline pion candidates are accumulated in the selected sample to obtain $N_{\text{Int}}$ and  $N_{\text{Inc}}$ in bins of kinetic energy.

Thus, the total inclusive interaction cross section can be measured, independent of the topology of the interaction, by taking the ratio of $N_{\text{Int}}$ to  $N_{\text{Inc}}$ for all pion-candidate tracks within the detector fiducial volume. In the next section, we describe how this method is applied in the analysis.

\subsection{Applying the ``Thin-Slice'' Method}

\begin{figure}
\includegraphics[width =\columnwidth ]{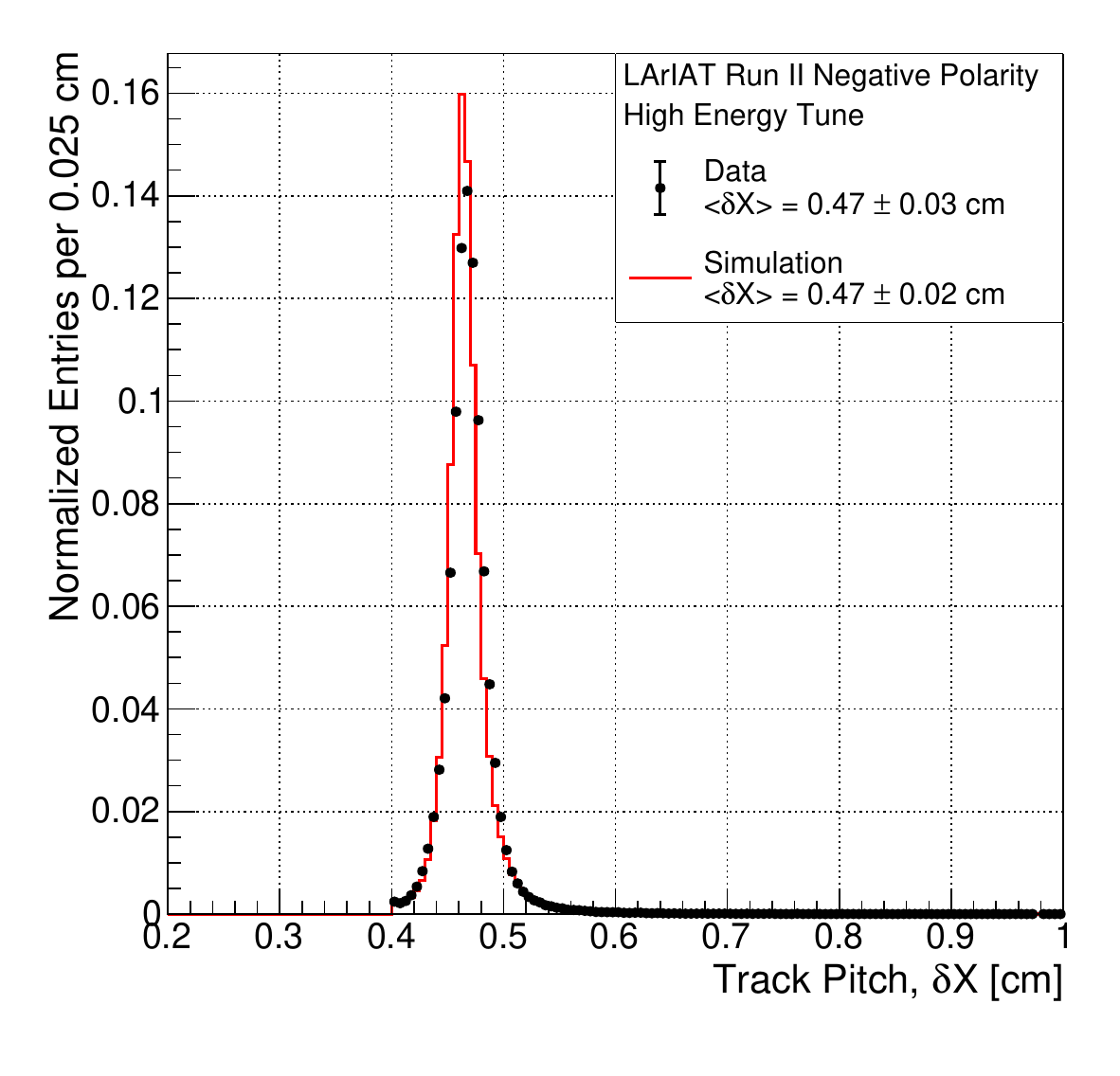}
\caption{\label{fig:pitch}  Pitch distribution for the Run~II negative polarity high energy tune data displayed in black, simulation in red, area normalized. Mean and standard deviation of the distributions are reported in the legend.
}
\end{figure}

\begin{figure}
\includegraphics[width =\columnwidth ]{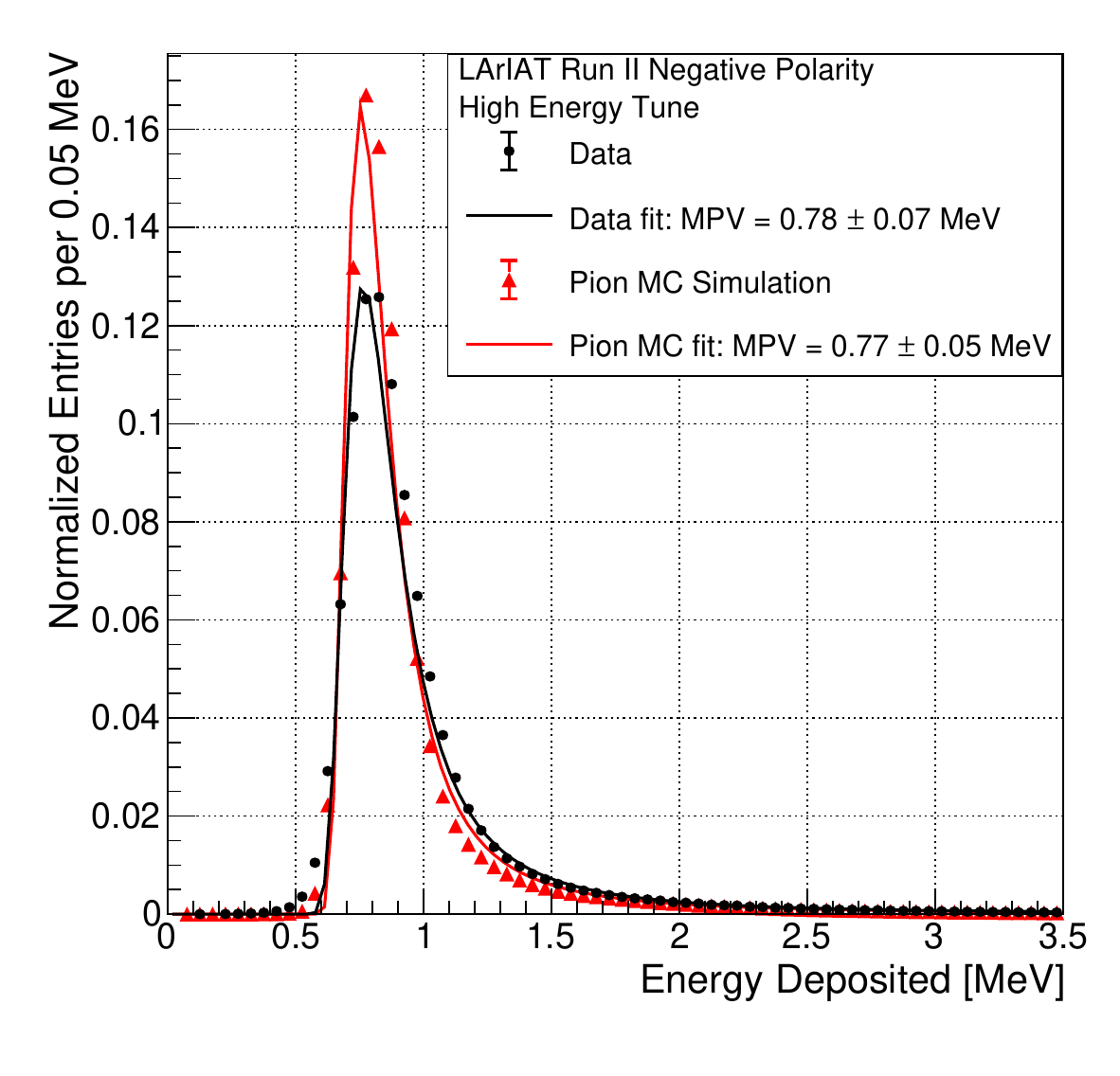}
\caption{\label{fig:enDep}  Per-slice energy deposition distribution for Run~II negative polarity high energy tune data displayed in black, simulation in red, with Landau fits to determine the most probable value and width of each distribution. The distributions are area normalized.}
\end{figure}

The tracking and calorimetry algorithms provide measurements of $E_{\text{Dep},n}$, the energy deposited along the pion's path at each segment between two 3D points of the trajectory, where the segment length is known as the ``track pitch'' ($\delta X$). Figure~\ref{fig:pitch} compares the distribution of track pitches for the high energy tune data and simulation, which are in agreement with a mean $\delta X$ of 4.7~mm and comparable widths (comparable agreement is found for the low energy tune data). Although the widths of these distributions are narrow, we assign to each slice a weight of ${\delta X}/{4.7}$, which allows us to simplify our machinery to use the fixed mean value of track pitch in the equations. The collection plane signals are used to obtain the measurement of deposited energy in each argon slice, shown in Figure~\ref{fig:enDep}. This per-slice energy deposition, together with the particle's momentum measurement from beamline instrumentation, allows us to assess the kinetic energy of the matched pion candidate at each point of the pion's reconstructed track.

\begin{figure}
\centering
\includegraphics[width=\columnwidth]{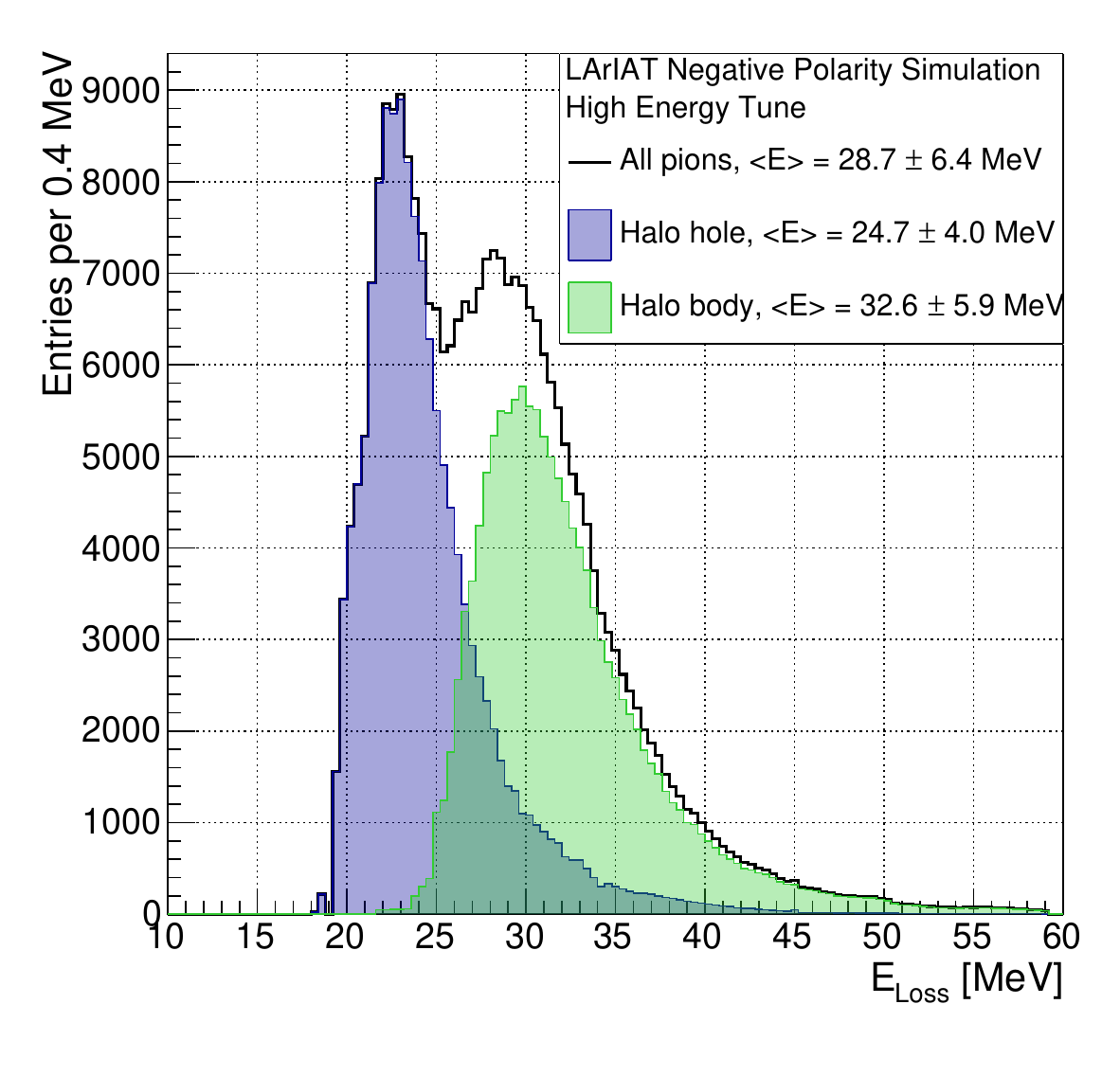}
\caption{\label{fig:ELoss100A}  True energy loss between the fourth wire chamber and the LArTPC front face according to the simulation of negative polarity events for the high energy tune. Two populations are visible, depending on the amount of material traversed by the beamline pion. The HALO scintillator paddle -- a rectangular piece with a central hole -- is responsible for the double-peaked distribution. The distribution for the pions traveling through the central hole (missing the scintillator material) is shaded in purple, and the distribution for the pions hitting the HALO body is shaded green, the sum of these distributions is indicated by the black histogram. }
\end{figure}

We start by estimating the pion's initial  kinetic energy as it enters the front face of the LArTPC, $ E^\text{kin}_{\text{Front Face}}$, as 
\begin{equation}
 E^\text{kin}_{\text{Front Face}}  =  \sqrt{p^2_{\text{Beam}} + m^2_{\text{Beam}}} - m_{\text{Beam}} - E_{\text{Loss}},
\label{eq:enFF}
\end{equation}
where $p_{\text{Beam}}$ is the momentum measured by the LArIAT spectrometer, the mass of the pion is $m_{\text{Beam}} = 139.57018\pm0.00035$~MeV~\cite{Patrignani:2016xqp}, $E_{\text{Loss}}$ is a correction for the kinetic energy lost in the uninstrumented material between the final wire chamber (WC4 in Figure~\ref{fig:beamlinebird}) and the LArTPC front face. Figure~\ref{fig:ELoss100A} shows the $E_{\text{Loss}}$ distribution for the simulation of pions in the LArIAT simulation. The lower of the two peaks is that of particles which pass through the central hole of the halo scintillator paddle, while the upper peak is particles that traverse the scintillator plastic body of the halo paddle, where they lose approximately 7~MeV more energy than the hole-crossing particles.

To obtain the kinetic energy at each point of the pion reconstructed track, we iteratively subtract the measurement of the energy deposited $E_{\text{Dep},n}$ from $ E^{kin}_{\text{Front Face}}$. This is shown in the application of the thin-slice method at the $j^{th}$ point of the track  $E_{j}^\text{kin}$ is given by
\begin{equation}
\begin{split}
 E_{j}^\text{kin}  & = E^{kin}_{\text{Front Face}} -  \sum_{n < j} E_{\text{Dep},n}.
\end{split}
\label{eq:KEj2}
\end{equation}


\subsection{\label{sec:Corrections0}Thin-Slice Flux Corrections}
In the evaluation of the interacting and incident fluxes as a function of the pion kinetic energy,  two corrections are necessary. The first accounts for beamline backgrounds. For this analysis these are the residual muons and electrons present in the beamline pion sample. The second correction accounts for LArTPC event reconstruction effects in the evaluation of  $N_{\text{Int}}$ and $N_{\text{Inc}}$.

We rewrite Eq.~\ref{eq:thinTargetXSSolved} at kinetic energy $E^{kin}_i$, as a function of the measured fluxes in data ($N^{\text{Data}}_{\text{Int}}$ and $N^{\text{Data}}_{\text{Inc}}$), with the corresponding corrections for beamline backgrounds ($C^{\pi \text{MC}}_{\text{Int}}$ and $C^{\pi \text{MC}}_{\text{Inc}}$), and corrections for reconstruction effects ($\psi^{\text{Int}}$ and $\psi^{\text{Inc}}$) as follows:
\begin{equation}
      \sigma^{\pi^-}_{TOT}(E^\text{kin}_{i})  = \frac{1}{n\text{ } \delta X}\frac{ \psi^{\text{Inc}}_i  \hspace{0.2cm} C^{\pi \text{MC}}_{\text{Int,i}}  \hspace{0.2cm} N^{\text{Data}}_{\text{Int,i}}  }{   \psi^{\text{Int}}_i \hspace{0.2cm} C^{\pi \text{MC}}_{\text{Inc,i}}  \hspace{0.2cm}  N^{\text{Data}}_{\text{Inc,i}} },
\label{eq:C}
\end{equation}
where the subscript $i$ highlights that the quantities are evaluated separately in each kinetic energy bin.

\subsubsection{Reconstruction Corrections ($\psi$)}
We rely on the tracking algorithm to identify the incoming pion in the LArTPC and its interaction point. The end point of the track within the fiducial volume determines whether or not the pion interacted.  The tracking algorithm has an intrinsic limitation in resolving shallow scattering angles, effectively placing a selection on the distribution of scattering angles included in the cross section measurement.  
Thus, we restrict the measured cross section to interaction angles greater than 5 degrees. 

The interacting and incident fluxes both depend on the tracking algorithm's ability to identify the interaction point, albeit in different ways. If the tracking does not stop at an interaction point and instead keeps adding subsequent hits to the pion trajectory (e.g., hits generated by an outgoing product of the interaction), the missed interaction constitutes an inefficiency for the interacting flux and the slices corresponding to the extra trajectory points result in over-counting for the incident flux. In contrast, if the tracking stops before the actual interaction point, then the identified interaction point constitutes an over-counting for the interaction flux at an energy higher than the true energy of the interaction, and the missed subsequent slices represent an inefficiency for the incident flux. We encode both the inefficiency and the over-counting due to the reconstruction of the pion track in the reconstruction corrections, $\psi^{\text{Int}}_i$ and $\psi^{\text{Inc}}_i$. We use a sample of DDMC pions to evaluate the reconstruction corrections as follows:
\begin{equation}
 \psi^{\text{Int}}_i  =  \frac{N^{\text{ Reco MC $\pi$}}_{\text{Int,i}}}{ N^{\text{ True MC  $\pi$}}_{\text{Int,i}}  } \hspace{0.5cm}\text{and}\hspace{0.5cm}  \psi^{\text{Inc}}_i  =  \frac{N^{\text{ Reco MC $\pi$}}_{\text{Inc,i}}}{ N^{\text{ True MC  $\pi$}}_{\text{Inc,i}}  },
\end{equation}
where $N^{\text{ True MC  $\pi$}}_{\text{Int,i}}$ is the expected number of interacting pions at the kinetic energy $E^{kin}_i$ according to the Geant4 10.03.p1 FTFP\_BERT interaction model and  $N^{\text{ Reco MC $\pi$}}_{\text{Int,i}}$ is the number of reconstructed interacting pions at the same kinetic energy. Analogous definitions apply to the correction on the incident flux. The values for $\psi^{\text{Inc}}_i/\psi^{\text{Int}}_i$ in each energy bin, and their corresponding statistical  and systematic uncertainties are reported in Table~\ref{tab:XSsummary}.

\subsubsection{Beamline Background Corrections ($C$)}
Pions are by far the largest component in the selected beamline candidates. However, a residual background of muons and electrons are present in the sample. We use simulations to estimate the pion content, $C^{\pi \text{MC}}_{\text{Int,i}}$ and  $C^{\pi \text{MC}}_{\text{Inc,i}}$, i.e., the percentage of pion entries in the interacting and incident flux at each kinetic energy bin, $E^{kin}_i$. We then apply these corrections to the corresponding kinetic energy bins in data.  

We simulate pions, muons, and electrons with the DDMC.  These are fed into the thin-slice method machinery, estimating the interacting and incident fluxes for each particle species. We then calculate the relative pion content for the interacting and incident flux as
\begin{equation}
C^{\pi \text{MC}}_{\text{Int,i}}  =  \frac{N^{\pi \text{MC}}_{\text{Int,i}}}{ N^{\text{TOT MC}}_{\text{Int,i}} } 
\end{equation}
and 
\begin{equation}
C^{\pi \text{MC}}_{\text{Inc,i}}  =  \frac{N^{\pi \text{MC}}_{\text{Inc,i}}}{ N^{\text{TOT MC}}_{\text{Inc,i}} } 
\end{equation}
where 
$N^{\text{TOT MC}}_{\text{Int,i}}$ and $N^{\text{TOT MC}}_{\text{Inc,i}}$ are the sum of the MC pion, muon, and electron contributions to the interacting and incident fluxes, respectively, while $N^{\pi}_{\text{Int,i}}$ and $N^{\pi}_{\text{Inc,i}}$ represent only the MC pion contribution. Once we evaluate the correction factors $C^{\pi \text{MC}}_{\text{Int,i}}$ and  $C^{\pi \text{MC}}_{\text{Inc,i}}$ separately, we use their ratio to background-correct the measured raw cross section. The values for $C^{\pi \text{MC}}_{\text{Int,i}}/C^{\pi \text{MC}}_{\text{Inc,i}}$ in each energy bin and statistical uncertainty are reported in Table \ref{tab:XSsummary}.

\begin{figure}
\centering
\includegraphics[width=\columnwidth]{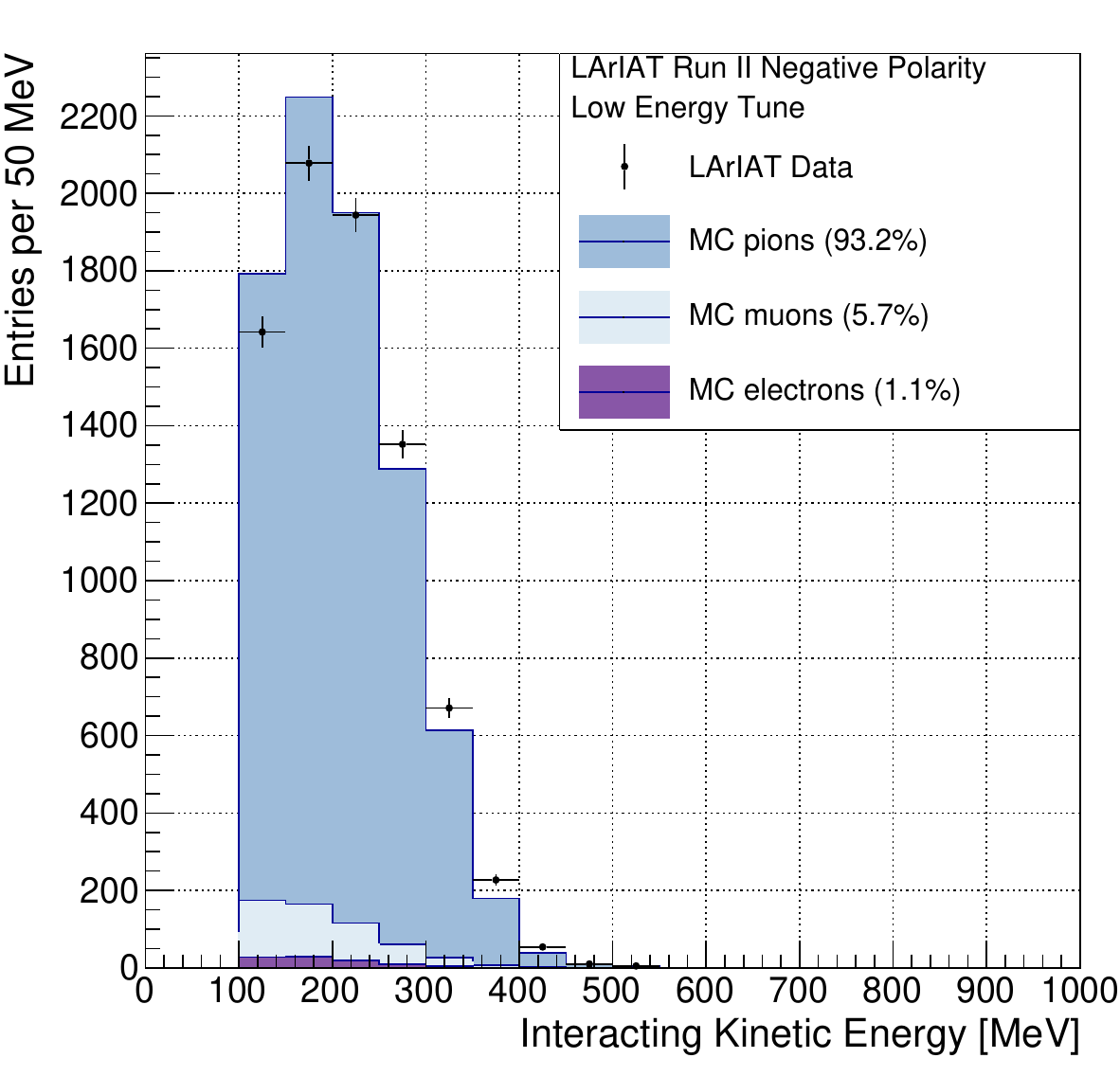}
\includegraphics[width=\columnwidth]{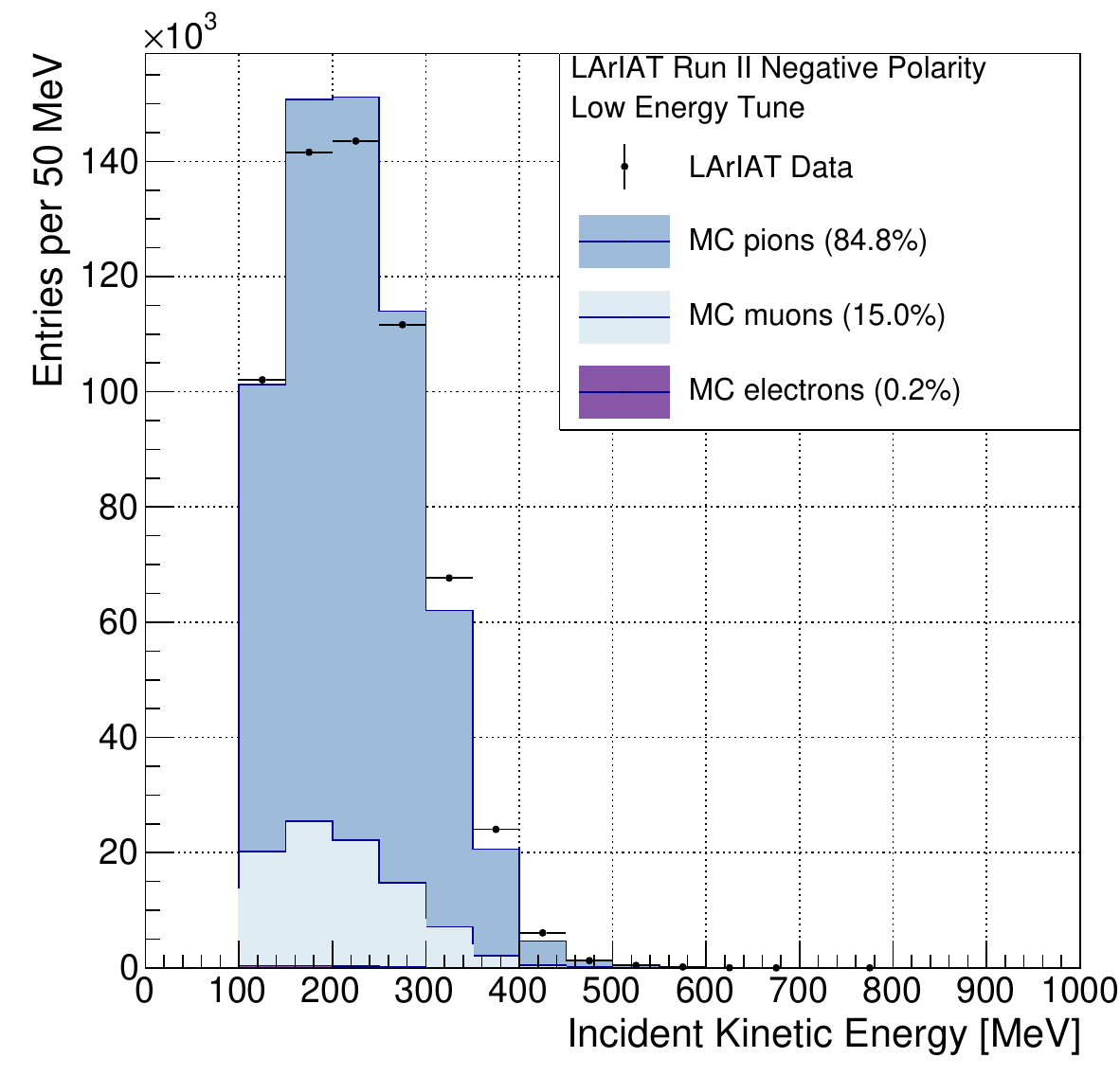}
\caption{\label{fig:60AFlux} Interacting and incident fluxes for the low energy tune. The simulated stacked contributions for pions (dark blue), muons (light blue) and electrons (purple) are normalized to the number of pion  candidates selected in low energy tune data.}
\end{figure}

\begin{figure}
\centering
\includegraphics[width=\columnwidth]{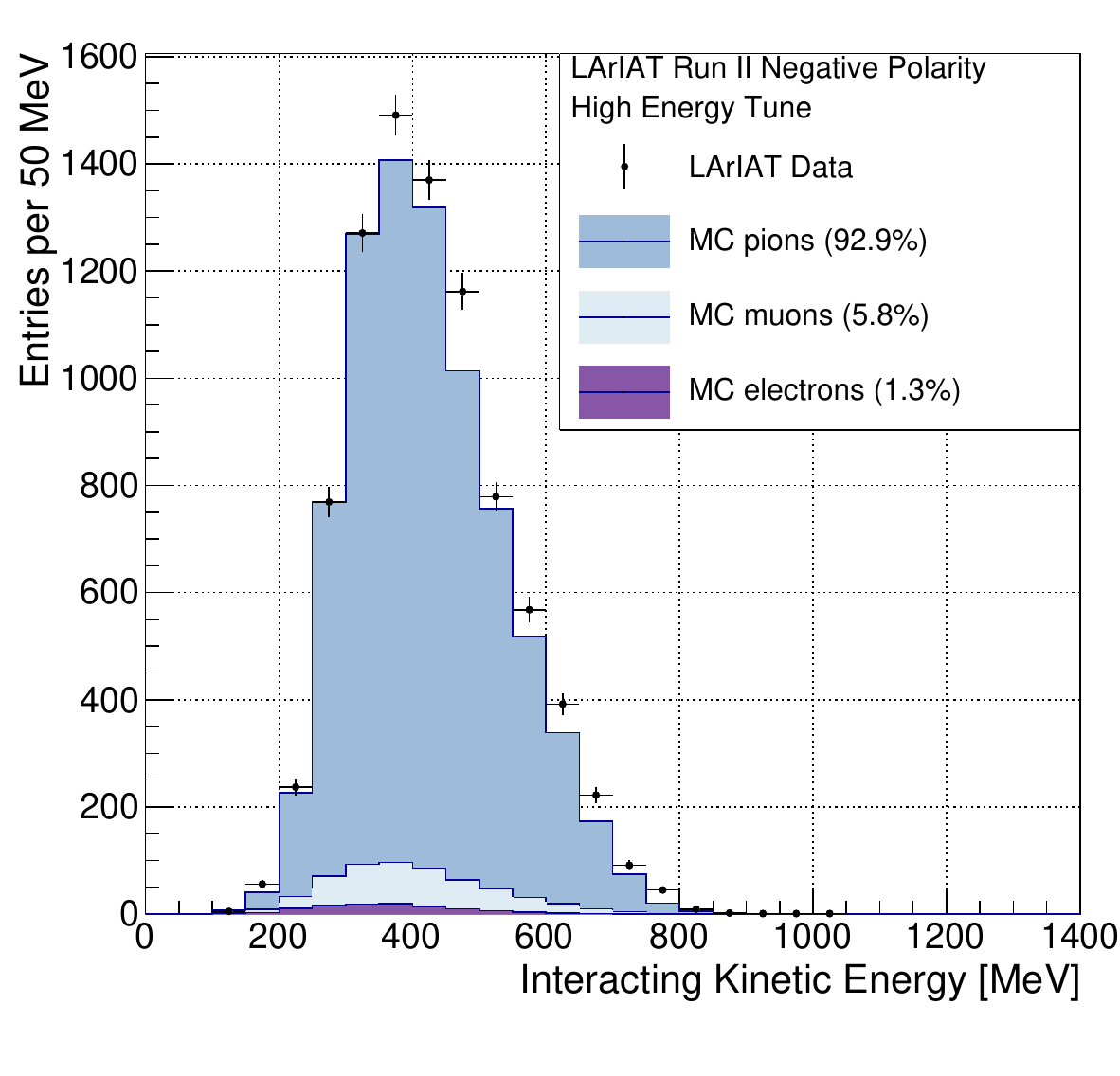}
\includegraphics[width=\columnwidth]{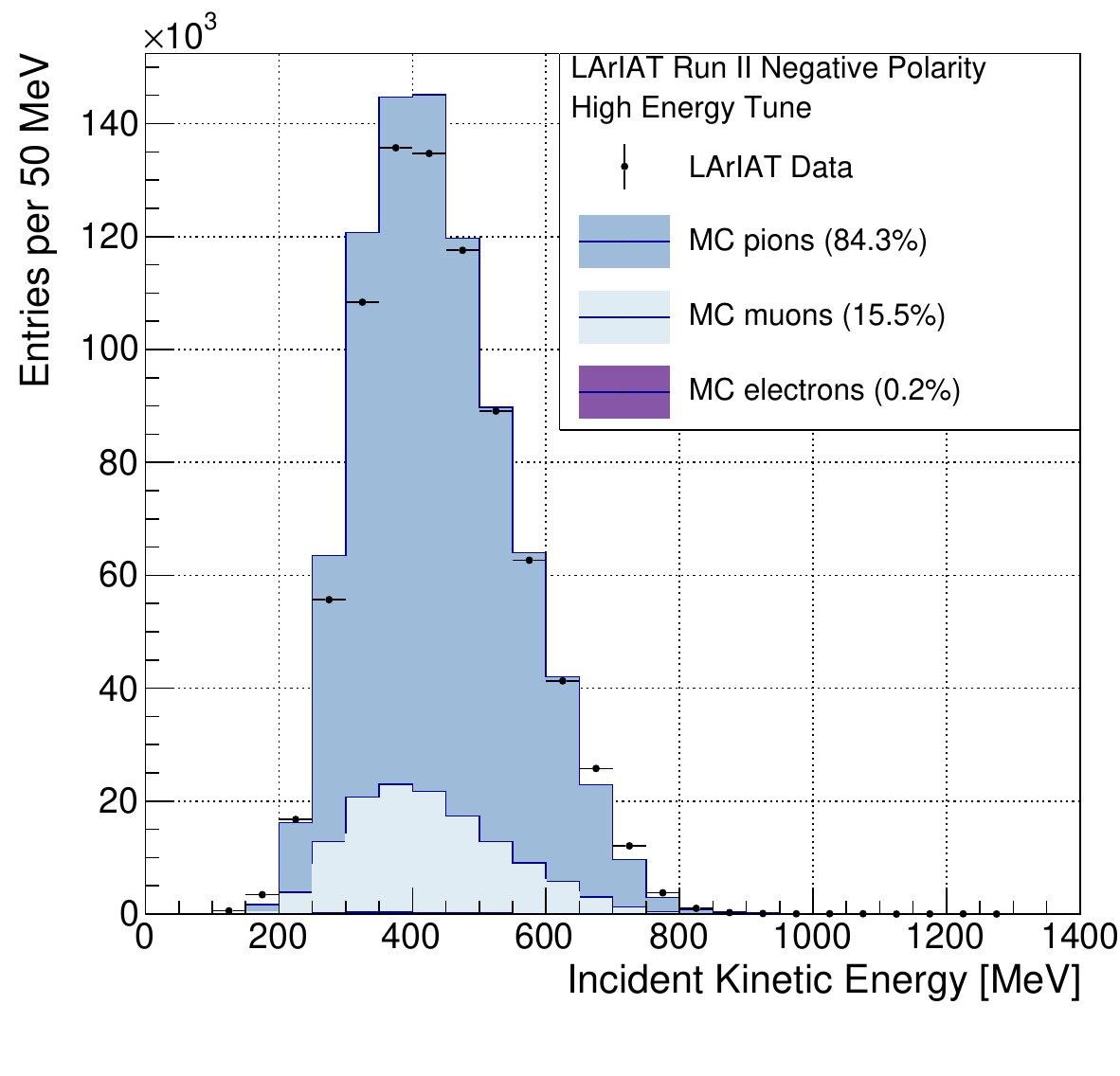}
\caption{\label{fig:100AFlux}  Interacting and incident fluxes for the high energy tune. The simulated stacked contributions for pions (dark blue), muons (light blue) and electrons (purple) are normalized to the number of pion  candidates selected in high energy tune data. }
\end{figure}

Muons represent the biggest contribution to beamline backgrounds. Without correction, the muon presence in the sample tends to lower the measured raw cross section. This is because most of the muons will cross the LArTPC without stopping, contributing almost exclusively to the incident flux. Figures~\ref{fig:60AFlux} and  \ref{fig:100AFlux} show the distributions of the interacting and incident fluxes as a function of the kinetic energy for the low and high energy tunes, respectively. While the agreement of data and simulation is reasonable, it should be noted that the Monte Carlo  prediction for pions assumes the Geant4 pion interaction cross section, which is not guaranteed to be correct. Indeed, these figures hint that our data will prefer a slightly different energy dependence of the cross section, which will be shown in the results.

\subsection{\label{sec:Corrections}Treatment of the uncertainties}
We calculate the statistical uncertainty for a given kinetic energy bin of the cross section by propagating the statistical uncertainty on $N_{\text{Int,i}}$ and $N_{\text{Inc,i}}$.  Since the number of incident particles in each energy bin is given by a simple counting, we assume that $N_{\text{Inc,i}}$ follows a Poisson distribution with mean and variance equal to $N_{\text{Inc,i}}$ in each bin.  
On the other hand, $N_{\text{Int,i}}$ follows a binomial distribution: a particle in a given energy bin might or might not interact.  The variance for the binomial is given by  
\begin{equation}
\text{\textsf{Var[}} N_{\text{Int,i}} \text{\textsf{]}}
 = N_{\text{Inc,i}}P_{\text{Int,i}}(1-P_{\text{Int,i}}),
\label{eq:binVar}
\end{equation}
where  $P_{\text{Int,i}}$ is estimated by $\frac{ N_{\text{Int,i}}}{N_{\text{Inc,i}}}$ and the number of tries corresponds to $N_{\text{Inc,i}}$. $N_{\text{Inc,i}}$ and $N_{\text{Int,i}}$ are not independent. In fact, populating a given bin for the interacting flux always implies at least populating the same bin in the incident flux  (and possibly other incident bins at higher energies). Thus, we conservatively calculate the statistical uncertainty on the cross section as 
\begin{equation}
\frac{\delta\sigma^{Stat}_{TOT}(E^\text{kin}_i)}{\sigma^{Stat}_{TOT}(E^\text{kin}_i)} =  \frac{\delta N_{\text{Int,i}}}{N_{\text{Int,i}}}+\frac{\delta N_{\text{Inc,i}}}{N_{\text{Inc,i}}}
\end{equation}
where:
\begin{eqnarray}
\delta N_{\text{Int,i}} &=& \sqrt[]{N_{\text{Int,i}}\Big(1-\frac{ N_{\text{Int,i}}}{N_{\text{Inc,i}}}\Big)}\\
\delta N_{\text{Inc,i}} &=& \sqrt[]{N_{\text{Inc,i}}} 
.
\end{eqnarray} 

Systematic uncertainties associated with the measured kinetic energy at each argon slice and with the beam composition are also evaluated and propagated through to the cross section.
The uncertainty on the kinetic energy of a pion candidate at the j$^{th}$ slice of argon  is given by

\begin{equation}
\delta E^\text{kin}_{j} = \sqrt{(\delta p_{\text{Beam}})^2 + (\delta E_{\text{Loss}})^2 +  ( \sum_{n < j} \delta E_{\text{Dep},n})^2}, 
\end{equation}
where $\delta p_{\text{Beam}}$ is the uncertainty associated with the beamline momentum measurement, $\delta E_{\text{Loss}}$ is the uncertainty associated with the energy loss in the uninstrumented material upstream of the LArTPC, and $\sum_{n < j}\delta E_{\text{Dep},n}$  is the uncertainty associated with pion energy deposition in each slice from the LArTPC front face to the $j^{th}$ slice. 

The momentum uncertainty,  $\delta p_{\text{Beam}}$, is estimated to be $2\%$ $p_{\text{Beam}}$, based on studies of multiple scattering and beamline geometry reported in Ref.~\cite{Acciarri_2020}.

The uncertainty in the amount of energy lost by particles traversing uninstrumented materials upstream of the LArTPC is estimated via the study illustrated in Figure~\ref{fig:ELoss100A}. The two populations visible in the figure are due to the path of the particle through the HALO scintillator paddle: particles that pass through the central hole of the paddle lose approximately 8~MeV less energy than particles that traverse scintillator plastic. The sum of those two populations is indicated by the black histogram, and we conservatively take its width, 6.4~MeV, as the uncertainty on this energy loss. 

We calculate the uncertainty associated with pion energy deposition at the j$^{th}$ slice as the sum of $\delta E_{\text{Dep},n}$ up to that slice; $\delta E_{\text{Dep},n}$ is 0.07~MeV,  which corresponds to the width of the distribution in  Figure~\ref{fig:enDep}. The measurements of the energy deposited in consecutive slices are not independent, because the charge collected on a given wire induces signal on neighboring wires and therefore contributes to the assessment of the energy deposited in neighboring slices. For this reason, a conservative simple sum of the uncertainties on the energy deposited in the single slice is employed when assessing the uncertainty on the total energy deposited by the pion.

We propagate the uncertainty on the kinetic energy to the cross section by varying the energy measurement, $E^\text{kin}_{j}$, at each argon slice and evaluating the cross section in three cases: first, we compute the cross section with the measured central value $E^\text{kin}_{j}$, then with an upward variation  $E^\text{kin}_{j} + \delta E^\text{kin}_{j}$, and finally with a downward variation  $E^\text{kin}_{j} - \delta E^\text{kin}_{j}$. The systematic uncertainty due to energy reconstruction is assigned for each bin as the maximum absolute value of excursion from the central value of that bin when the cross section is calculated for the two cases: ($E^\text{kin}_{j} + \delta E^\text{kin}_{j}$) and ($E^\text{kin}_{j} - \delta E^\text{kin}_{j}$).

We assess the systematic uncertainty on the beam composition in a similar fashion; we vary the beamline muon and electron content  by the respective uncertainty as reported in Table~\ref{tab:MCafterCutContaminants}, and evaluate the cross section variation in each case.

The systematic uncertainty on the $\psi$ corrections is evaluated by varying the underlying elastic and inelastic cross sections in our simulation. 
The elastic scattering contribution to the total cross section was adjusted based on the difference between the two elastic scattering models available in Geant4: G4Elastic and G4DiffuseElastic. In each kinetic energy bin, we conservatively used as the variation the maximum difference between those models, weighted by the angular distribution of the events in that bin. The corresponding variations for the inelastic model were derived under the assumption that the overall total cross section normalization is accurate at the few percent level.
With these shape and normalization variations applied to the underlying cross sections, the systematic uncertainty on the ratio $\psi_{Inc}/\psi_{Int}$ in each bin was taken as the magnitude of the excursion from the ratio obtained using the nominal cross section model. These uncertainties are reported in Table~\ref{tab:XSsummary}.

\section{\label{sec:Results0}Results}
\begin{figure}[b]
\includegraphics[width =\columnwidth ]{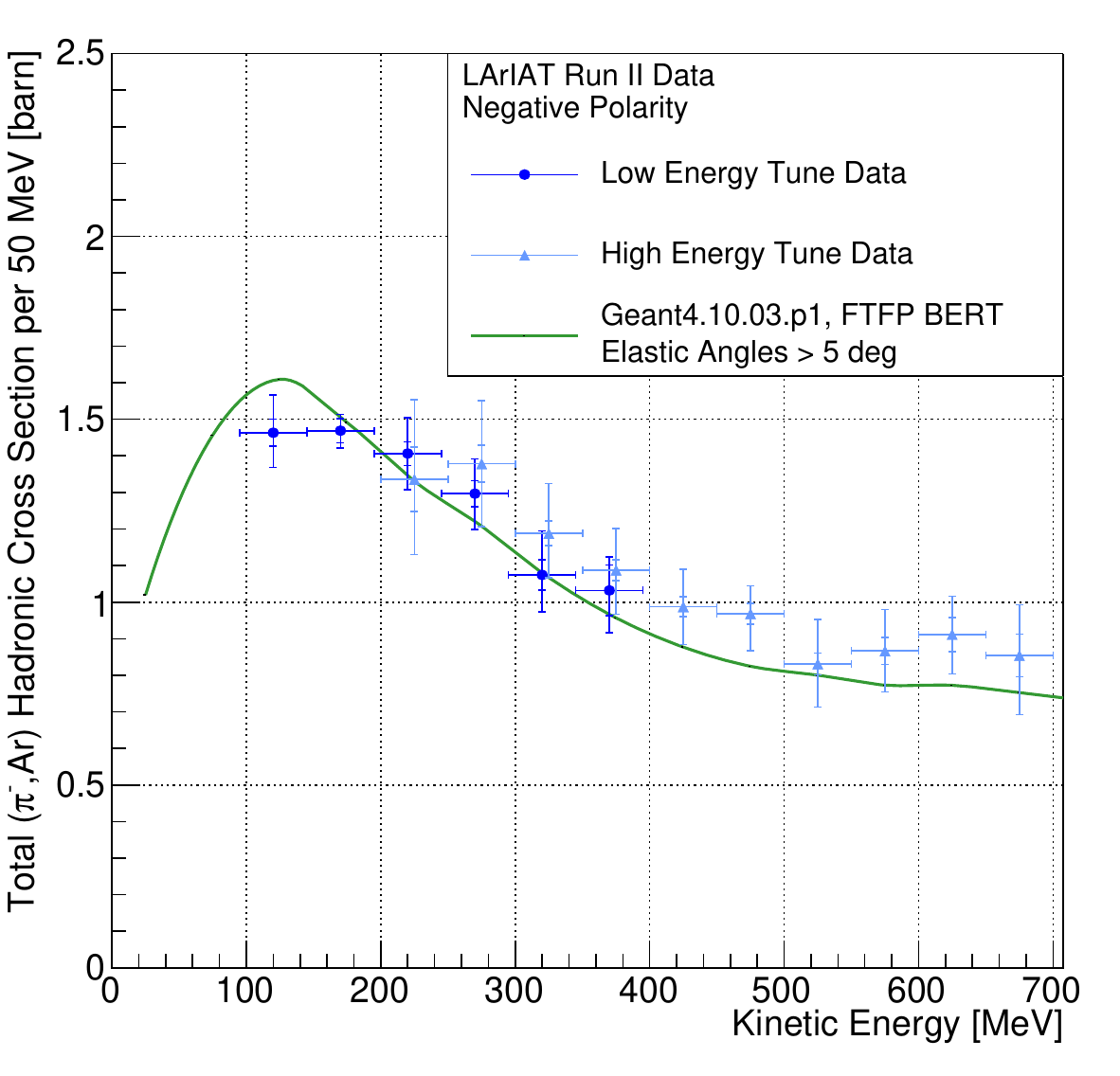}
\caption{\label{fig:epsart} ($\pi^-$, Ar) total hadronic cross section for  scattering angles greater than 5~degrees measured in the LArIAT low energy and high energy tune Run~II data samples. The Geant4 prediction for the total hadronic cross section for inelastic scattering and elastic scattering angles greater than 5~degrees is displayed in green. }
\end{figure}

Figure~\ref{fig:epsart} shows the measured ($\pi^-$, Ar) total hadronic cross section,  for  scattering angles greater than 5~degrees, as calculated according to Eq.~\ref{eq:C} in the 100-700~MeV kinetic energy range. Table~\ref{tab:XSsummary} summarizes the cross section as a function of kinetic energy and  the associated total uncertainty; it also reports the recorded fluxes,  beamline background corrections, and the  corrections due to pion track reconstruction effects. The lower bound of the kinetic energy range was chosen to eliminate physical backgrounds due to non-hadronic interactions such as pion decay at rest and pion capture. Since both of these processes occur at rest, a kinetic energy cut-off of 100~MeV completely eliminates them.

Our experimental result is presented separately for the high energy and low energy tune data, which are consistent with each other.  
For each dataset, the lowest and highest bins reported are those that contained at least 100 interactions. We report both the statistical uncertainty (inner ticks) and the sum in quadrature of statistical and systematic uncertainty (outer ticks). For reference, the black curve represents the Geant4 prediction based on the FTFP\_BERT cascade model  for inelastic scattering and elastic scattering angles greater than 5~degrees. 
In most bins,  statistical uncertainty dominates. Since the low statistics mostly result from the hard LArTPC fiducialization selection, improvements in the track reconstruction algorithm in future analyses will have the potential to recover these events.
The dominant systematic uncertainty results from our uncertainty on the muon background. 

Data from the high and low energy tunes was independently collected and simulated with dedicated simulated samples. We applied identical methodology and selection criteria to each tune. The total cross section measurement for the two samples presents a sizable overlap in the kinetic energy. In the overlapping energy region (200-400~MeV), the measured values from the two data sets agree within statistical uncertainty, providing an internal validation for the analysis method.

The measurement shows the typical shape caused by the underlying $\Delta$ resonance production, as expected from pion interactions below 500~MeV.
Our measurement is found to be in general agreement with the expectation from the Geant4 FTFP\_BERT cascade model in the 100-700~MeV region, but with possible differences in the energy dependence of the cross section as compared with the prediction. In the higher energy region (above 400~MeV), the experimental measurement of the ($\pi^-$, Ar) cross section is slightly higher than the prediction. This result represents the first dedicated measurement of the negative pion hadronic interaction cross section on an argon target.

\begin{table*}[p]
\caption{\label{tab:XSsummary} Results summary, including slice counts and corrections for backgrounds and reconstruction effects.} 
\begin{ruledtabular}
\begin{tabular}{cccccccc}
$[E^\text{kin}_{\text{MIN}}, E^\text{kin}_{\text{MAX}}]$& $\sigma$ Geant4  & $\sigma_{\text{TOT}}$ & Stat $\bigoplus$ Syst  & $N^{ \text{Data}}_{ \text{Int}}$
& $N^{ \text{Data}}_{ \text{Inc}}$ & $C^{\pi \text{MC}}_{\text{Int}} / C^{\pi \text{MC}}_{\text{Inc}}$ & $ \psi^{\text{Inc}}/\psi^{\text{Int}}$ \\ 
(MeV)& (Barn)& (Barn) & Uncertainty (Barn) & & & &  \\\hline
& & & & & & & \\
Low E Tune& & & & & & & \\
& & & & & & & \\
$[ 100. , 150. ]$& 1.61  & 1.46 &+0.10/-0.10 & 1642 & 102070 & $1.13 \pm 0.03$ & $0.80 \pm 0.05$\\
$[ 150. , 200. ]$& 1.49  & 1.47 &+0.04/-0.05 & 2078 & 141574 & $1.11 \pm 0.02$ & $0.89 \pm 0.02$\\
$[ 200. , 250. ]$& 1.33  & 1.41 &+0.10/-0.10 & 1944 & 143522 & $1.10 \pm 0.03$ & $0.93 \pm 0.05$\\
$[ 250. , 300. ]$& 1.21  & 1.30 &+0.10/-0.10 & 1352 & 111641 & $1.09 \pm 0.03$ & $0.97 \pm 0.06$\\
$[ 300. , 350. ]$& 1.07  & 1.07 &+0.12/-0.10 & 671  & 67668  & $1.08 \pm 0.05$ & $0.99 \pm 0.04$\\
$[ 350. , 400. ]$& 0.96  & 1.03 &+0.09/-0.12 & 227  & 24041  & $1.07 \pm 0.08$ & $1.01 \pm 0.02$\\
& & & & & & & \\
\hline
& & & & & & & \\
High E Tune & & & & & & & \\
& & & & & & & \\
$[ 200. , 250. ]$& 1.33  & 1.34 &+0.22/-0.21 &  237 & 16760  & $1.12 \pm 0.08$ & $0.83 \pm 0.13$ \\
$[ 250. , 300. ]$& 1.21  & 1.38 &+0.17/-0.17 &  769 & 55680  & $1.14 \pm 0.04$ & $0.87 \pm 0.09$ \\
$[ 300. , 350. ]$& 1.07  & 1.19 &+0.14/-0.12 & 1271 & 108371 & $1.12 \pm 0.03$ & $0.90 \pm 0.08$ \\
$[ 350. , 400. ]$& 0.96  & 1.09 &+0.11/-0.12 & 1491 & 135718 & $1.11 \pm 0.03$ & $0.88 \pm 0.08$ \\
$[ 400. , 450. ]$& 0.88  & 0.99 &+0.10/-0.10 & 1370 & 134715 & $1.10 \pm 0.03$ & $0.87 \pm 0.08$ \\
$[ 450. , 500. ]$& 0.82  & 0.98 &+0.08/-0.10 & 1162 & 117560 & $1.10 \pm 0.04$ & $0.88 \pm 0.07$ \\
$[ 500. , 550. ]$& 0.80  & 0.83 &+0.12/-0.12 & 779  & 89095  & $1.09 \pm 0.04$ & $0.86 \pm 0.12$ \\
$[ 550. , 600. ]$& 0.77  & 0.87 &+0.11/-0.11 & 568  & 62669  & $1.10 \pm 0.05$ & $0.86 \pm 0.11$ \\
$[ 600. , 650. ]$& 0.75  & 0.91 &+0.10/-0.11 & 392  & 41301  & $1.10 \pm 0.06$ & $0.87 \pm 0.11$ \\
$[ 650. , 700. ]$& 0.73  & 0.85 &+0.14/-0.16 & 222  & 25790  & $1.09 \pm 0.08$ & $0.90 \pm 0.13$ \\
\end{tabular}
\end{ruledtabular}
\end{table*}

\section{\label{sec:Results}Acknowledgements}
This document was prepared by the LArIAT collaboration using the resources of the Fermi National Accelerator Laboratory (Fermilab), a U.S. Department of Energy, Office of Science, HEP User Facility. Fermilab is managed by Fermi Research Alliance, LLC (FRA), acting under Contract No. DE-AC02-07CH11359. We also gratefully acknowledge the support of the National Science Foundation; Brazil CNPq grant number 233511/2014-8, Coordena\c{c}\~ao de Aperfei\c{c}oamento de Pessoal de N\'ivel Superior - Brazil (CAPES) - Finance Code 001, S\~ao Paulo Research Foundation - FAPESP (BR) grant number 16/22738-0; Polish National Science Centre grant Dec-2013/09/N/ST2/02793; the Science and Technology Facilities Council (STFC), part of the United Kingdom Research and Innovation; The Royal Society (United Kingdom); and the JSPS grant-in-aid (Grant Number 25105008), Japan. The collaboration extends a special thank you to the coordinators and technicians of the Fermilab Test Beam Facility, without whom none of this work would have been possible.


\bibliography{bib.bib}

\end{document}